%% file: CTQMC_Manuscript.tex
\documentclass[preprint,11pt]{elsarticle}

\usepackage{amsmath}
\usepackage{bera}
\usepackage{amssymb}
\usepackage{graphicx}
\usepackage{xcolor, listings}
\usepackage{url}
\usepackage{geometry}
\usepackage{bm}

\newcounter{bla}

\colorlet{punct}{red!60!black}
\definecolor{background}{HTML}{EEEEEE}
\definecolor{delim}{RGB}{20,105,176}
\colorlet{numb}{magenta!60!black} 
 
\lstdefinelanguage{json}{
    basicstyle=\small\ttfamily,
    numbers=left,
    numberstyle=\scriptsize,
    stepnumber=1,
    numbersep=8pt,
    showstringspaces=false,
    breaklines=true,
    frame=lines,
    backgroundcolor=\color{background},
    literate=
     *{0}{{{\color{numb}0}}}{1}
      {1}{{{\color{numb}1}}}{1}
      {2}{{{\color{numb}2}}}{1}
      {3}{{{\color{numb}3}}}{1}
      {4}{{{\color{numb}4}}}{1}
      {5}{{{\color{numb}5}}}{1}
      {6}{{{\color{numb}6}}}{1}
      {7}{{{\color{numb}7}}}{1}
      {8}{{{\color{numb}8}}}{1}
      {9}{{{\color{numb}9}}}{1}
      {:}{{{\color{punct}{:}}}}{1}
      {,}{{{\color{punct}{,}}}}{1}
      {\{}{{{\color{delim}{\{}}}}{1}
      {\}}{{{\color{delim}{\}}}}}{1}
      {[}{{{\color{delim}{[}}}}{1}
      {]}{{{\color{delim}{]}}}}{1},
}

 \newcommand{\ctqmcname}{ComCTQMC{}} 

\newcommand*{\citen}{}
\DeclareRobustCommand*{\citen}[1]{%
  \begingroup
    \romannumeral-`\x 
    \setcitestyle{numbers}%
    \cite{#1}%
  \endgroup
}

\journal{Computer Physics Communications}

\begin{document}

\begin{frontmatter}

\title{Accelerated impurity solver for DMFT and its diagrammatic extensions}

\author[b]{Corey Melnick\corref{author}}
\author[a]{Patrick S\'{e}mon}
\author[a]{Kwangmin Yu}
\author[a]{Nicholas D'Imperio}
\author[c]{Andr\'{e}-Marie Tremblay}
\author[a,d]{Gabriel Kotliar}

\cortext[author] {Corresponding author.\\\textit{E-mail address:} cmelnick@bnl.gov}
\address[a]{Brookhaven National Laboratory,
Computational Science Initiative,
Upton, NY }
\address[b]{Brookhaven National Laboratory,
Department of Condensed Matter Physics and Materials Science,
Upton, NY }
\address[c]{Universit\'{e} de Sherbrooke, 
D\'{e}partement de physique and Regroupement qu\'{e}b\'{e}quois sur les mat\'{e}riaux de pointe,
Sherbrooke, Qu\'{e}bec, Canada }
\address[d]{Rutgers University,
Department of Physics,
Piscataway, NJ}

\begin{abstract}
We present \ctqmcname{}, a GPU accelerated quantum impurity solver. It uses the continuous-time quantum Monte Carlo (CTQMC) algorithm wherein the partition function is expanded in terms of the hybridisation function (CT-HYB). \ctqmcname{} supports both partition and worm-space measurements, and it uses improved estimators and the reduced density matrix to improve observable  measurements whenever possible. \ctqmcname{} efficiently measures all one and two-particle Green's functions, all static observables which commute with the local Hamiltonian, and the occupation of each impurity orbital.  \ctqmcname{} can solve complex-valued impurities with crystal fields that are hybridized to both fermionic and bosonic baths. Most importantly, \ctqmcname{} utilizes graphical processing units (GPUs), if available, to dramatically accelerate the CTQMC algorithm when the Hilbert space is sufficiently large. We demonstrate acceleration by a factor of over 600 (100) in a simulation of $\delta$-Pu at 600 K with (without) crystal fields. In easier problems, the GPU offers less impressive acceleration or even decelerates the CTQMC.  Here we describe the theory, algorithms, and structure used by \ctqmcname{} in order to achieve this set of features and level of acceleration.

\end{abstract}


\end{frontmatter}

\newpage

{\bf PROGRAM SUMMARY}

\begin{small}
\noindent
{\em Program Title:}  \ctqmcname \\
{\em Licensing provisions(please choose one): } GPLv3 \\
{\em Programming language:} C++/CUDA \\
{\em Supplementary material:}  \\
{\em Nature of problem:} In dynamical mean-field theory (DMFT), the computational bottleneck is the repeated solution of a quantum impurity problem [1]. The  continuous-time quantum Monte-Carlo (CTQMC) algorithm has emerged as one of the most efficient methods for solving multiorbital impurity problems at moderate-to-high temperatures [2]. However, the low-temperature  regime remains inaccessible, particularly for $f$-shell systems, and the measurement of two-particle correlation functions on an impurity adds a substantial computational burden. The bottleneck of the CTQMC solver is itself the computation of the local trace which includes the multiplication of many moderate-to-large sized matrices. The efficient solution of the impurity, measurement of the two-particle correlation functions, and acceleration of the  trace computation are therefore critical.     \\
{\em Solution method:}  \ctqmcname{}  uses the hybridisation expansion of the impurity action to explore partition space [3]. It uses the worm algorithm [4] to explore the union of the partition space with observables spaces, e.g., the two-particle correlation functions. It uses improved estimators to more accurately measure the one- and two-particle Green's functions  [5]. Identical impurities are solved across all MPI ranks (for ideal weak scaling) and the trace computations of these impurities are distributed to and accelerated by GPUs (when available). The lazy-trace algorithm [6] is used to further reduce the burden of the local trace calculation.   \\
{\em Additional comments including Restrictions and Unusual features:} \ctqmcname{} solves nearly arbitrary impurities, including those with complex valued and time-dependent interactions. However, there are two restrictions: (1) The retarded part of the interaction is described by a set of bilinears (a paired creation and annihilation operator), and these bilinears must commute with the local Hamiltonian and have real quantum numbers; (2) If a local Green's function vanishes, then the corresponding hybridisation function also vanishes.\\

\end{small}

\section{Introduction}
\label{sec:Introduction}
\input{Src/Introduction}

\section{Theory}\label{sec:Theory}
\input{Src/Theory}

\section{Implementation}\label{sec:Implementation}
\input{Src/Implementation}


\section{Examples and results}\label{sec:Examples}

\input{Src/Examples}

\section{Summary}
 
Here we have presented the program \ctqmcname{} which efficiently solves Anderson impurity problems using the CTQMC algorithm with worm and/or partition space sampling. This program utilizes GPUs extremely well in the most difficult problems. Indeed, we demonstrate the acceleration of the CTQMC algorithm by a factor of over 600 in the simulation of a low-symmetry $f$-shell impurity. The parallelisation and acceleration schemes are extremely flexible, and it can in general use 100\% of the computational capacity of HPC resources to solve these difficult problems. The algorithm is similarly flexible: It can handle complex valued impurities with dynamical interactions and crystal fields; and it can use worm sampling to measure any non-zero component of a two-, three-, or four-point susceptibility in the particle-hole or particle-particle channels. Finally, it is highly optimized, using state-of-the-art algorithm and improved estimators to drastically reduce the time-to-solution.  

Additionally, equations for the improved estimators in the presence off a dynamical interaction are presented. Details of the implementation are discussed, and examples are provided. Best-practices and convergence criteria are discussed in these examples. Appendices are also provided to derive equations or investigate conclusions discussed in \ctqmcname{}. A user guide is also provided with \ctqmcname{} which discusses, among other topics, its installation and usage.
 
\section{Acknowledgments}
This work was supported by the US Department of Energy, Office of Basic Energy Sciences as part of the Computation Material Science Program. This research used resources of the Oak Ridge Leadership Computing Facility, which is a DOE Office of Science User Facility supported under Contract DE-AC05-00OR22725.

\bibliographystyle{elsarticle-num}
\bibliography{CTQMC_Manuscript.bib}

\appendix
\input{Src/Appendices}

\end{document}

%% file: Src/Introduction.tex
Here we introduce \ctqmcname, a GPU accelerated quantum impurity solver which uses the continuous-time quantum Monte Carlo (CTQMC) algorithm wherein the impurity action is expanded in orders of the hybridisation function (CT-HYB). \ctqmcname{} can compute the one- and two-particle vertex functions, susceptibilities, and Green's functions of an impurity; it can measure the value of static observables which commute with the local Hamiltonian or good quantum numbers of the local Hamiltonian; and it can provide energy-dependent valence histograms for such observables. In combination with its post-processing tools, \ctqmcname{} provides the ability to carry out dynamical mean-field theory (DMFT)\cite{Kotliar1996,Kotliar2006} for both models or real materials (in, e.g., the LDA or GW approximations).
Indeed, \ctqmcname{} has been embedded in the open source, strongly-correlated physics packages ComSuite\cite{Choi2019} and Portobello\cite{Portobello}. 

DMFT plays a crucial role in the study of strongly-correlated materials wherein density function theory (DFT) fails to accurately predict even the most basic material properties. Where DFT presumes that the electronic structure can be described via Landau Fermi-liquid theory, i.e., as itinerant quasiparticles, DMFT is a minimal model which treats both itinerant and localized electrons on an equal footing.\cite{Kotliar1996,Kotliar2006}  In DMFT, one maps the correlated physics of a system onto a local impurity which hybridizes with a non-interacting bath of particles. The Weiss field acting upon the impurity due to this bath is determined by the self-consistent solution of the DMFT equations, which requires the repeated solution of an Anderson impurity model. Indeed, the fast and accurate solution of the Anderson impurity models lies at the heart of any efficient DMFT implementation. 

To address this need, a number of impurity solvers have been developed. These include the exact diagonalization (ED)\cite{Kotliar1996,Schuler2015}, numerical renormalization group (NRG)\cite{Bulla2008}, and quantum Monte-Carlo (QMC)\cite{Gull2011} methods.  While the QMC algorithms cannot generally access very low temperatures and can encounter a sign-problem after which the impurity becomes practically unsolvable, QMC methods have become the most widely used impurity solver largely because they are flexible and numerically exact. That is, a QMC algorithm can be applied to problems with arbitrary interactions and couplings, phases, and symmetries; and the error vanishes systematically as the computational time is increased.\cite{Gull2011}

In the QMC algorithm, a Markov chain is generated by sampling the diagrams of the (imaginary time) impurity action via the Metropolis Hastings algorithm. That is, operators are stochastically placed on the imaginary time axis according to the relative weight of the new diagram as compared to the current diagram. In the CTQMC algorithm, these operators are drawn anywhere on the imaginary time axis. (This method contrasts with the original Hirsch and Fry algorithm\cite{Hirsch1986} which discretizes the imaginary time axis and suffers from systematic error as a result.) In order to efficiently compute the relevant weights of each diagram in partition space, one must expand the action in terms of some part of the impurity Hamiltonian. Common choices include the interaction (CT-INT), an auxiliary field (CT-AUX), or the hybridisation between bath and impurity (CT-HYB).\cite{Gull2011} CT-HYB has become the most common choice for solving the impurity model in real materials, as it is the most capable of solving general multiorbital problems at low temperatures.

While CT-HYB is one of the best methods for solving multiorbital impurity problems, the computational burden associated with an $f$-shell system or a difficult cellular DMFT (CDMFT) problem is immense, particularly at low temperatures. Compounding this problem is the strong desire in the strong-coupling community to compute the susceptibilities of strongly correlated materials or to diagrammatically extend DMFT using the two-particle response functions. Indeed, the two-particle correlation functions require substantially more computational resources to accurately estimate than the one-particle objects required by DMFT. As a result of the massive computational hurdles facing DMFT and its extensions, a veritable host of algorithms have been developed to enhance the ability of CT-HYB to quickly sample the partition function (lazy trace computation\cite{semon2014lazy}, binary trees\cite{Gull2008}, basis truncations\cite{Shim2007}, Krylov implementations\cite{Lauchli2009}); to efficiently store two-particle objects and filter the high-frequency noise\cite{Boehnke2011,Shinoaka2017b}; to more efficiently sample many one- and two-particle objects (improved estimators\cite{Hafermann2012,Gunacker2016}); to directly sample configuration spaces which are rarely or never explored by the partition function (worm algorithm\cite{Gunacker2015,Gunacker2016}); and to reconstruct the high-frequency regions of the self-energy and one-particle Green's function\cite{Gull2011} or full vertex\cite{Kaufmann2017} using their asymptotic forms. Nearly all of these aforementioned improvements are included in the \ctqmcname{} package. Where conflicts arise, we try to use the most recent and optimal algorithms.

Other packages, e.g., TRIQS\cite{seth2016triqs}, iQIST\cite{Huang2015}, and ALPSCore CT-HYB\cite{Shinoaka2017a}, have implemented their own set of highly optimized algorithms. However, \ctqmcname{} can leverage a powerful and emerging feature of modern supercomputers: the graphical processing unit (GPU). While essentially all CTQMC codes are massively parallelizable across computer processing units (CPU), none of the major CTQMC codes have utilized GPUs. As we will show, this is a massive advantage when solving the most difficult problems. 

A typical compute node on a modern supercomputer has $O(10)$ CPU cores distributed between a few physical CPUs. While the CPUs have access to a large amount of memory, only a very small subset of this memory is on the physical CPU. In comparison, a GPU designed for high-performance computing (HPC) has $O(1000)$ cores and drastically more on-device memory. The NVIDIA Volta used on the Summit supercomputer at Oak Ridge National Laboratory, for example, has 2,560 cores capable of handling double precision computations, each of which has access to 16 GB of shared memory across the device and 9 MB in an L2 cache, which is orders of magnitude larger than any CPU. While these cores are not as fast or as good at handling branching logic trees as a CPU, their ability to process large volumes of data and parallelize computation across this data is unsurpassed. Therefore, a GPU can perform many (non-branching) operations on a large object much faster than a CPU could; and if small and distinct portions of an algorithm require such a computation, these kernels can be handed from CPU to GPU in order to accelerate the algorithm. Provided that these computations represent the bottleneck of the calculation, the performance gain can be substantial. 

The quintessential example of a suitable kernel is in the multiplication of two rank $n$ matrices, which requires $n^3$ operations. If $n$ is sufficiently large, a GPU can compute the result thousands of times faster than a CPU. As we will discuss, such matrix multiplications form a bottleneck in CT-HYB, and GPUs can therefore be used to dramatically accelerate the algorithm.

\begin{figure}
\centering
\includegraphics[scale=1]{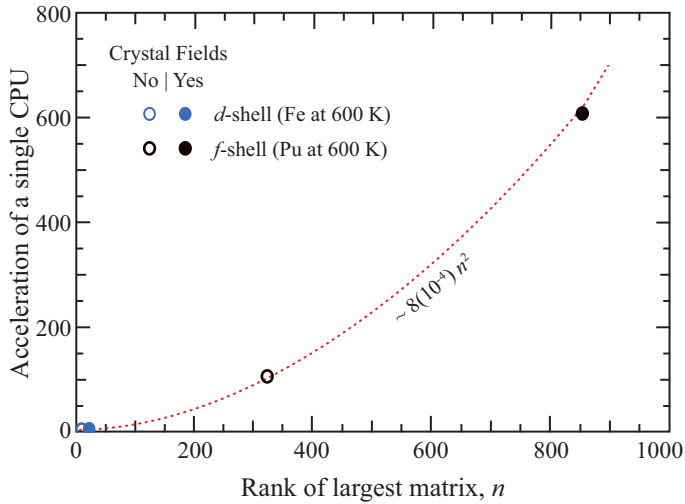}
\caption{Acceleration of the CTQMC algorithm on a single CPU by a single GPU on Summit at ORNL for variations in the problem size. The problem size is described here by the rank of the largest matrix handled by the algorithm, which depends on the size and symmetries of the Hilbert space. The larger the Hilbert space and the fewer the symmetries, the larger the matrices become, the harder the problem becomes, and the more acceleration a GPU offers. Here we show the acceleration of representative $d$ and $f$-shell problems (which have Hilbert spaces of size $2^n=1,024$ and $65,536$, respectively, where $n=10$ and 14 are the number of correlated orbitals) with and without crystal field effects (which reduce the symmetries of the problem).}
\label{fig:gpu_acceleration_1}
\end{figure}

In CTQMC, the bottleneck in computation occurs each time one updates configuration of the Markov chain (and recomputes the trace over the local impurity states). This requires the multiplication of many moderate-to-large sized matrices. As we have discussed, this is precisely the type of problem for which a GPU was designed. In our implementation, however, it is not only the matrix multiplications which are handled on a GPU: our GPU kernels are used to compute the matrix norm, which is crucial in the lazy-trace algorithm\cite{semon2014lazy}, apply the time-evolution operators, add matrices, and compute the final trace. Not only does this allow the GPU to handle more computations, but it also nearly eliminates the costly transfer of the matrices between CPU and GPU memory. Furthermore, we have the GPU accelerate multiple simulations simultaneously in order to more fully utilize its capabilities. With this approach, we show that a GPU can accelerate the time-to-solution of a single CPU by a factor of over 600 on Summit!  Furthermore, our implementation is flexible: even for relatively small problems, e.g., where the largest matrices have a rank of only $n_{max}=26$, some small GPU acceleration is achieved (5 times), as shown in Fig. \ref{fig:gpu_acceleration_1}. Still, if the problem is sufficiently small, GPU-CPU communication will hurt performance: the leftmost data-point in Fig. \ref{fig:gpu_acceleration_1} is decelerated by a factor of 10. In Sec. \ref{sec:trace}, we will discuss the use of GPUs in much greater detail. Appendix \ref{app:Summit} provides the relevant  details of our deployment on Summit and physical approximations which lead to these results. (The associated input files are included with the distribution of \ctqmcname{}.)

Before getting into the implementation, it is useful to overview the theory behind\ctqmcname{} and discuss its capabilities and limitations.
In Sec. \ref{sec:Theory} we present the theory behind the CTQMC algorithm, i.e., the quantum impurity model and its action; the algorithm used to sample partition space and the modifications used to sample worm spaces;  and the observables one can measure in these spaces and the improved estimators used to more accurately and efficiently measure one- and two-particle vertices (e.g., the self-energy). Then, we will discuss the implementation in Sec. \ref{sec:Implementation}, i.e., we will describe the structure of the code, its optimization, and its acceleration using GPUs. Finally, we present a few results in Sec. \ref{sec:Examples}, discussing the computational burden involved in measuring various quantities and some best-practices for measuring them. 

A guide for installation, tutorials with input files and converged results, and tables describing all input parameters are included with \ctqmcname{}. In the appendices, we provide supporting information. In particular, we define the basis functions used in \ctqmcname{} which are used for the simulation of real materials ({\ref{app:basis}), a description of the Metropolis-Hastings algorithm  (\ref{sec:metropolis-hasting}), and a discussion of impurities where partition space sampling fails (\ref{app:EDimpurities}). We also provide a description of the optimized moves used in \ctqmcname{} (\ref{sec:CTQMCMoveOptimizations}), and a description of the symmetries  enforced on two-particle correlators in order to more quickly converge these observables (\ref{app:susceptibilities}). 

Now, let us overview the theory behind the CT-HYB algorithm and its worm-sampling extension.

%% file: Src/Theory.tex
In this section, we briefly discuss the theoretical foundation of  \ctqmcname.  That is, we discuss the general quantum impurity model solved by \ctqmcname, the hybridisation expansion of the action, and the resulting CTQMC algorithm used to sample partition space, the worm algorithm used to sample observable spaces, and the improved estimators used to more accurately measure variouos observables. Finally, we discuss the various observables which can be computed by \ctqmcname. 

\subsection{Quantum impurity model}\label{sec:Quantum Impurity Model}

An impurity model consists of a small interacting system, the impurity, which hybridizes with baths of non-interacting particles. We consider here hybridization with both a fermionic and a bosonic bath, and the different contributions to the impurity model Hamiltonian, $\hat{H}$, are split into the purely local part, $\hat{H}_{\text{loc}}$; the fermionic and bosonic parts, $\hat{H}_{\text{bath},f}$ and $\hat{H}_{\text{bath},b}$; and the hybridisation between the the baths and the impurity, $\hat{H}_{\text{hyb},f}$ and $\hat{H}_{\text{hyb},b}$. That is,
\begin{equation}
\label{equ:ImpurityModelHamiltonian}
\hat{H} = \hat{H}_{\text{loc}} + \hat{H}_{\text{hyb},f}  + \hat{H}_{\text{bath},f} + \hat{H}_{\text{hyb},b} + \hat{H}_{\text{bath},b},
\end{equation}
The local part of the impurity Hamiltonian includes both one- and two-body interactions. That is, 
\begin{equation}
\label{equ:ImpurityHamiltonian}
\hat{H}_\text{loc} = \sum_{ij} \hat{c}_i^\dagger (t_{ij} - \mu \delta_{ij})\hat{c}_j + \frac{1}{2}\sum_{ijkl} \hat{c}_i^\dagger \hat{c}_j^\dagger U_{ijkl} \hat{c}_k\hat{c}_l,
\end{equation}
where $c^\dagger_i$ creates a fermion in the generalized orbital $i$, $t_{ij}$ are hopping amplitudes, $U_{ijkl}$ is the interaction tensor, and $\mu$ is the chemical potential.
The hybridization with the fermionic bath is described by 
\begin{equation}
\hat{H}_{\text{hyb},f} = \sum_{i\lambda} \hat{c}_i^\dagger V_{i\lambda} \hat{f}_\lambda + \hat{f}_\lambda^\dagger V^*_{i\lambda} \hat{c}_i  \quad\quad\text{and} \quad\quad  \hat{H}_{\text{bath},f} = \sum_\lambda \epsilon_\lambda \hat{f}_\lambda^\dagger \hat{f}_\lambda,
\end{equation}
where $\hat{f}_\lambda^\dagger$ creates a fermion in the bath orbital $\lambda$ with energy $\epsilon_\lambda$, and the $V_{i\lambda}$ are fermionic hybridization amplitudes. The hybridization with the bosonic bath is described by 
\begin{equation}
\hat{H}_{\text{hyb},b} = \sum_{I\kappa}  W_{I\kappa}(\hat{b}_\kappa + \hat{b}_\kappa^\dagger) \hat{Q}_I   \quad\quad\text{and} \quad\quad  \hat{H}_{\text{bath},b} = \sum_\kappa \omega_\kappa \hat{b}_\kappa^\dagger \hat{b}_\kappa,  
\end{equation}
where $\hat{b}_\kappa^\dagger$ creates a boson in the bath orbital $\kappa$ with energy $\omega_\kappa$, and the $W_{I\kappa}$ are hybridization amplitudes between the bosonic bath and charge degrees of freedom on the impurity, 
\begin{equation}
\label{equ:ParticleHoleDef}
\hat{Q}_I = \sum_{ij}  \langle I  |  ij \rangle \hat{c}_i^\dagger \hat{c}_j.
\end{equation}
These particle-hole bilinears $\hat{Q}_I$ are Hermitian, and the hybridization amplitudes $W_{I\kappa}$ are real.    

The path integral formalism allows us to integrate out the bath degrees of freedom, and the action of the impurity model can be written as
\begin{align}
\label{equ:Action}
S = &-\sum_{ij} \iint_0^\beta \overline{c}_i(\tau) \mathcal{G}_{ij}^{-1}(\tau - \tau') c_j(\tau') d\tau d\tau'  \notag\\
&+ \frac{1}{2}\sum_{ijkl} \iint_0^\beta \overline{c}_i(\tau^+) \overline{c}_j(\tau'^+) \mathcal{U}_{ijkl}(\tau - \tau') c_k(\tau')c_l(\tau) d\tau d\tau'.
\end{align}
The the Weiss field  
\begin{equation}
\label{equ:WeissField}
\mathcal{G}_{ij}^{-1}(i\nu_n) = (i\nu_n + \mu)\delta_{ij} - t_{ij} - \Delta_{ij}(i\nu_n) 
\end{equation}
absorbs the fermionic bath degrees of freedom, which are encapsulated in the hybridization function 
\begin{equation}
\label{equ:hybridization}
\Delta_{ij}(i\nu_n) = \sum_\lambda \frac{V_{i\lambda}V_{j\lambda}^*}{i\nu_n - \epsilon_\lambda}.
\end{equation}
Similarly, the dynamic interaction 
\begin{equation}
\label{equ:Interaction}
\mathcal{U}_{ijkl}(i\omega_n) = U_{ijkl} + \sum_{IJ}\langle kj | I \rangle D_{IJ}(i\omega_n)\langle J | il  \rangle
\end{equation}
absorbs the bosonic bath degrees of freedom, which are encapsulated in the bosonic hybridisation function
\begin{equation}
D_{IJ}(i\omega_n) = \sum_{\kappa} \frac{2W_{I\kappa}W_{J\kappa}\epsilon_\kappa}{(i\omega_n)^2 - \epsilon_\kappa^2}.
\end{equation}
Note that we denote the fermionic Matsubara frequencies by $i\nu_n$ and the bosonic Matsubara frequencies by $i\omega_n$. Recall as well that the hybridisation amplitudes, $W_{I\kappa}$, are real. 

\ctqmcname{} can solve most impurities which can be defined within this formalism: It  supports complex valued hopping amplitudes $t_{ij}$, interaction tensors $U_{ijkl}$, and hybridisation functions (i.e., impurity models where not all $V_{i\lambda}$ can be chosen real, or equivalently, if $\Delta_{ij}(i\nu_n)\ne \Delta_{ji}(i\nu_n)$). Such complex valued impurities may arise when applying a complex valued unitary transformation to the one-particle basis used by default in CTQMC. (See \ref{app:basis}.) However, there are a few restrictions:
\begin{enumerate}
\item The particle-hole bilinears $\hat Q_{I}$ in Eq.~\ref{equ:ParticleHoleDef} need to satisfy
\begin{equation}
\label{equ:ParticleHoleCondition}
[\hat{H}_\text{loc}, \hat{Q}_I] = 0 \quad \quad \text{and} \quad \quad  [\hat{c}_i, \hat{Q}_I] = q_{iI} \hat{c}_i.
\end{equation}
Notice that, as a consequence, the $\hat{Q}$'s commute among each other and that the quantum numbers $q_{iI}$ are real. This is because the last equation is equivalent to requiring that $\hat{Q}_I = \sum_i q_{iI}\hat{c}_i^\dagger \hat{c}_i$ and that the particle-hole bilinears are Hermitian.  
\item The second restriction concerns the block-diagonal shape of the hybridization function, $\Delta_{ij}$, and the local Green's function matrix, $G_{\text{loc},ij}(\tau)=-\langle c_i(\tau)\overline{c}_j \rangle_\text{loc}$, where $\langle \circ \rangle_\text{loc}$ denotes the thermal average with respect to the impurity Hamiltonian $\hat{H}_\text{loc}$. The requirement is
\begin{equation}
G_{\text{loc},ij} \equiv 0  \quad \quad \Rightarrow  \quad \quad \Delta_{ij} \equiv 0.
\end{equation}  
Equivalently, the non-zero blocks of the hybridization function must lie within the non-zero blocks of the atomic Green function. Since the block-diagonal shape of $G_\text{loc}$ is usually a consequence of the abelian symmetries of $H_\text{loc}$, we may also say that the hybridization is not allowed to break the abelian symmetries of the impurity. 
\end{enumerate}

Having defined the impurity model and its restrictions, let us write the problem in a form amenable to Monte-Carlo sampling.

\subsection{Expansion}\label{sec:Expansion}

The CT-HYB algorithm begins by expanding the partition function of the impurity model in powers of the fermionic and bosonic hybridization functions. Using Eq.~\ref{equ:ParticleHoleCondition} to integrate out the bosonic part of the expansion, this yields
\begin{equation}
\label{equ:Zexpansion}
Z=\int \mathcal{D}[\overline{c}, c] e^{-S} = \sum_{k=0}^\infty \int_0^\beta \! d\tau_1  \cdots \int^{\tau_{k-1}}_0  \!\!\! d\tau_k \int_0^\beta \! d\tau_1' \cdots \int^{\tau_{k-1}'}_0 \!\!\! d\tau_k' \sum_{i_1\cdots i_k} \sum_{i_1'\cdots i_k'} w(\mathcal{C})
\end{equation} 
where $\mathcal{C} =  i_1\tau_1i_1'\tau_1' \cdots i_k\tau_k i_k'\tau'_k$ is the current configuration. The weight  of this configuration, $w(\mathcal{C})$, is the product of three distinct terms:  the local weight, the fermionic bath weight, and the bosonic bath weight. 
This decomposition, $w(\mathcal{C})=w_\text{loc}(\mathcal{C})w_\text{hyb}(\mathcal{C})w_\text{ret}(\mathcal{C})$, will prove convenient in the future.
These weights are given by the following equations:
\begin{align}
w_\text{loc}(\mathcal{C}) &= \mathrm{Tr}e^{-\beta \tilde{H}_{\text{loc}}} T_\tau \prod_{r=1}^{|\mathcal C |} \hat{c}_{i_r'}(\tau_r')\hat{c}_{i_r}^\dagger(\tau_r), \label{equ:pweight1} \\
w_\text{hyb}(\mathcal{C}) &= \underset{1\leq r,s\leq |\mathcal C |}{\mathrm{Det}} \Delta_{i_ri_s'}(\tau_r-\tau'_s), \label{equ:pweight2} \\
\text{and} \quad\quad\quad w_\text{ret}(\mathcal{C}) &= \exp \biggl ( \frac{1}{2}\sum_{r,s=1}^{2|\mathcal C |} \sum_{IJ}  \tilde{q}_{r I}\tilde{q}_{s J}K_{IJ}(\tilde{\tau}_r - \tilde{\tau}_s) \biggr ), \label{equ:pweight3}
\end{align}
respectively, where $|C|$ is the expansion order, i.e., the number of operator pairs in the trace. In Eq.~\ref{equ:pweight1}, the effective local Hamiltonian is defined by  
\begin{equation}
\label{equ:effectiveHamiltonian}
\tilde{H}_\text{loc} = \hat{H}_\text{loc} + \frac{1}{2}\sum_{IJ} \hat{Q}_I D_{IJ}(i\omega_n = 0) \hat{Q}_J
\end{equation}
and $\hat{c}_i(\tau) = e^{\tau \tilde{H}_\text{loc}} \hat{c}_i e^{-\tau \tilde{H}_\text{loc}}$. In Eq.~\ref{equ:pweight3}, the imaginary times of the configuration are relabeled as $\tilde \tau_r = \tau_r$ and $\tilde \tau_{r+|\mathcal C |} = \tau_r'$ with $r=1,\dots,|\mathcal C |$, and the quantum numbers are relabeled as $\tilde{q}_{rI} = q_{i_r I}$ and $\tilde{q}_{(r+|\mathcal C|)I} = -q_{i_r' I}$. Finally, 
\begin{equation}
\label{equ:DynSecondPrimitive}
K_{IJ}(\tau) = -\frac{1}{2\beta}D_{IJ}(i\omega_n = 0) |\tau|(\beta - |\tau|)+ \sum_{n \ne 0} \frac{D_{IJ}(i\omega_n)}{\beta(i\omega_n)^2}e^{-i\omega_n\tau}
\end{equation}
is the $\beta$-periodic second primitive of $D_{IJ}(\tau) - D_{IJ}(i\omega_n = 0)\delta(\tau)$.

Just as we have written an expansion for the partition function, we can also write an expansion for a local observable, $\mathcal{O}$,
\begin{equation}
\label{equ:Oexpansion}
\int \mathcal{D}[\overline{c}, c] e^{-S} \mathcal{O} = \sum_{k=0}^\infty \int_0^\beta \! d\tau_1  \cdots \int^{\tau_{k-1}}_0  \!\!\! d\tau_k \int_0^\beta \! d\tau_1' \cdots \int^{\tau_{k-1}'}_0 \!\!\! d\tau_k' \sum_{i_1\cdots i_k} \sum_{i_1'\cdots i_k'} w(\mathcal{C}, \mathcal{O}),
\end{equation} 
where the weights $w(\mathcal{C}, \mathcal{O})$ are again the product of three distinct terms. Assuming that the observable is of the form 
\begin{equation}
\label{equ:ObservableQuantumNumbers}
\mathcal O = O_1 (\tau_1'')\cdots O_l( \tau_l'') \quad \quad \text{with}\quad\quad [\hat O_a, \hat{Q}_I]=q_{aI}'\hat O_{a},
\end{equation}
the three terms are
\begin{align}
w_\text{loc}(\mathcal{C}, \mathcal{O}) &= \mathrm{Tr}e^{-\beta \tilde{H}_{\text{loc}}} T_\tau \hat O_{1}(\tau_1'')\cdots \hat O_{l}(\tau_l'')\prod_{r=1}^{|\mathcal C|} \hat{c}_{i_n'}(\tau_r')\hat{c}_{i_r}^\dagger(\tau_r), \\
w_\text{hyb}(\mathcal{C}, \mathcal{O}) &= \underset{1\leq r,s\leq |\mathcal C |}{\mathrm{Det}} \Delta_{i_ri_s'}(\tau_r-\tau'_s), \\
\text{and} \quad\quad\quad w_\text{ret}(\mathcal{C}, \mathcal{O}) &= \exp \biggl ( \frac{1}{2}\sum_{r,s=1}^{2|\mathcal C | + |\mathcal O|} \sum_{IJ}  \tilde q_{r I} \tilde q_{s J}K_{IJ}(\tilde \tau_r - \tilde \tau_s) \biggr ),
\end{align}
where $|\mathcal O|$ describes the number of operators in the observable.
Here $\tilde \tau_r$ and $\tilde{q}_{rI}$ are as defined in Eq.~\ref{equ:pweight3} for $r=1,\dots,2|\mathcal C|$, and for $r=2|\mathcal C | + 1,\dots,2|\mathcal C | + |\mathcal O|$ we define $\tilde \tau_{2|\mathcal C | + a} = \tau_a''$ and $\tilde q_{(2|\mathcal C | + a)I}=q_{aI}'$.
Notice that the contribution from the fermionic hybridization is the same as for the partition function expansion; the local operators of the observable do not hybridize with the fermionic bath.

The expansions of the partition function, Eq.~\ref{equ:Zexpansion}, and an observable,  Eq.~\ref{equ:Oexpansion}, establish the prerequisites for estimating observables within CTQMC.  As we will discuss in the following sections, there are two classes of observables: observables for which we only need to sample the partition function configuration space, and observables for which we must sample both the partition space and the observable space.

\subsection{Partition-CTQMC}
In this section, we consider observables which can be estimated by sampling the partition space. The expectation value of these observables can be written as the expectation value of a random variable over the partition function expansion,  
\begin{equation}
\label{equ:PartitionOExpansion}
\langle \mathcal O \rangle = Z^{-1}\sum_\mathcal{C} w(\mathcal C) o(\mathcal C, \mathcal O) =\langle o(\mathcal C, \mathcal O) \rangle_{w(\mathcal C)/Z},
\end{equation}
where $o(\mathcal C, \mathcal O)$ is the random variable and $w(\mathcal C)/Z$ is the probability distribution. The quantity $o(\mathcal C, \mathcal O)$ is also called the ``estimator'' of the observable $\mathcal O$ in the configuration $\mathcal C$. We assume that $w(\mathcal C)$ is a positive real number for the time being. We will come back to the general case where $w(\mathcal C)$ is negative or complex in Sec.~\ref{sec:PartitionObservables}, but before doing so, let us briefly discuss the sampling of the partition function expansion.

\subsubsection{Sampling} \label{sec:PartitionSampling}
The probability distribution $w(\mathcal C)/Z$ is sampled by a Markov chain $\mathcal C_1 \rightarrow \mathcal C_2 \rightarrow \dots$ in the space of configurations, characterized by the transition probability $P(\mathcal{C}_{i + 1}|\mathcal{C}_i)$ of going from configuration $\mathcal C_i$ to configuration $\mathcal C_{i + 1}$. The Markov process converges to $w(\mathcal C)/Z$ if the transition probability satisfies the detailed balance $P(\mathcal C_{i+1}|\mathcal C_i)w(\mathcal C_i) = P(\mathcal C_i| \mathcal C_{i+1})w(\mathcal C_{i+1})$ and ergodicity conditions. 

The Metropolis-Hasting algorithm (\ref{sec:metropolis-hasting}) gives a possible choice for the transition probability. To start, a trial configuration $\mathcal C$ is chosen according to a trial probability $q(\mathcal C| \mathcal C_i)$, and we set $\mathcal C_{i+1}:=\mathcal C$ with probability  
\begin{equation}
p=\text{min}\biggl (\frac{q(\mathcal C_i| \mathcal C)}{q(\mathcal C |\mathcal C_i)}\times \frac{w(\mathcal C)}{w(\mathcal C_i)},\,\, 1 \biggr)
\end{equation}
and $\mathcal C_{i + 1} := \mathcal C_i$ otherwise. This transition probability $p\cdot q$ satisfies detailed balance.

For an ergodic sampling of the configuration space, one must propose updates (with an associated transition probability) that allow the Markov chain to explore all of configuration space. A natural choice is to propose the insertion or removal of a pair of operators. Assuming that the fermionic hybridisation does not break the abelian symmetries of the impurity, this choice is generally ergodic. If this assumption does not hold, one must insert four or more operators at once in order to ensure ergodicity, as discussed in Ref.~\cite{SemonErgodicity2014} (where the hybridization breaks charge conservation) and Ref.~\cite{seth2016triqs} (where the hybridization breaks orbital parity conservation). Additional updates, or moves, are also required when sampling observable spaces, as discussed in Ref. \cite{Gunacker2016}.

 By default, \ctqmcname{} only considers pair insertion and removal when exploring partition space, although the user may specify that four operator insertion and removal should also be considered. Additional moves are considered in observable spaces, as we will discuss. For additional detail on the moves implemented in \ctqmcname{}, e.g., the values of $q(\mathcal{C}_i|\mathcal{C})$, we direct the reader to \ref{sec:CTQMCMoveOptimizations}.

\subsubsection{Estimators} \label{sec:PartitionObservables}

The expectation value of an observable, $\mathcal O$, may be cast as the expectation value of a random variable over the partition space. For example, one can use Eq.~\ref{equ:Zexpansion} and Eq.~\ref{equ:Oexpansion} to write
\begin{equation}
\label{equ:OExpansionSimple}
\langle \mathcal O \rangle  =  Z^{-1} \sum_\mathcal{C} w(\mathcal C, \mathcal O) = Z^{-1} \sum_\mathcal{C} w(\mathcal C) \frac{w(\mathcal C, \mathcal O)}{w(\mathcal C)} , 
\end{equation}
which yields the estimator $w(\mathcal C, \mathcal O)/w(\mathcal C)$ for observable $\mathcal O$. In Eq.~\ref{equ:OExpansionSimple} we assume that $w(\mathcal C, \mathcal O)$ vanishes whenever $w(\mathcal C)$ vanishes. Let us start with observables for which this generally holds.   

First, let us discuss the \textbf{reduced density matrix} of the impurity, $\hat \rho$. This observable allows us to calculate the expectation value of any static observable $\hat{O}$ on the impurity. That is, $\langle \hat O \rangle = \mathrm{Tr}\hat \rho \hat{O}$. The estimator for the reduced density matrix is 
\begin{equation}
\label{equ:ImpurityDensityMatrix}
(\hat{\rho})_{uv} = \beta^{-1}\biggl \langle \int_0^\beta \frac{ w_\text{loc}(\mathcal C, \hat{P}_{vu}(\tau))}{w_\text{loc}(\mathcal C)}d\tau \biggr \rangle_{w(\mathcal C)/Z}.
\end{equation}
Here $\hat{P}_{vu}=|v\rangle \langle u |$, where $|u\rangle$ and $|v\rangle$ are states of the local (impurity) Hilbert space. In this estimator, we take advantage of time translational invariance and integrate over $\tau$ in order to reduce statistical noise. (This integration is done analytically.) Note that in this estimator there is no contribution from the bosonic bath, or equivalently, $w_\text{ret}$. This follows from the restriction that the hybridisation is not allowed to break the abelian symmetries of the impurity, such that $(\hat{\rho})_{uv}=0$ if $[\hat{P}_{vu},\hat{Q}_I]\ne 0$.  Notice that without this restriction, there generally are non-zero entries in impurity reduced density matrix which cannot be calculated by partition function sampling.    

By default, \ctqmcname{} will use the reduced density matrix to calculate the average \textbf{occupation} of each orbital, the average \textbf{energy} of the impurity, and the average \textbf{number} of electrons on the impurity. Users may supply additional quantum numbers or observables which may be computed using the reduced density matrix. CTQMC will check during post-processing that these observables meet the criterion: i.e., the quantum numbers are, indeed, quantum numbers of the local Hilbert  space, and the observables commute with the local Hamiltonian. It will keep those observables and quantum numbers which are acceptable and discard those which  are not. 

Next, consider the \textbf{susceptibility of a quantum number},  $\chi_{IJ}(\tau) = \langle Q_I(\tau) Q_J \rangle  - \langle Q_I \rangle \langle Q_J \rangle$.
If the particle-hole bilinears satisfy Eq.~\ref{equ:ParticleHoleCondition}, the estimator reads
\begin{equation}
\chi_{IJ}(i\omega_n) = \frac{1}{\beta (i\omega_n)^2}\biggr \langle  \sum_{r,s=1}^{2|\mathcal C |} \sum_{IJ}  \tilde{q}_{s I}\tilde{q}_{r J}e^{i\omega_n(\tilde{\tau}_s - \tilde{\tau}_r)} \biggr \rangle_{w(\mathcal C)/Z}.
\end{equation}
In \ctqmcname{}, we measure only the dynamical parts, $i\omega_n \ne 0$, as the static measurement is better estimated from the reduced density matrix, and the Fourier transform is taken analytically. (The $\tilde q$'s and $\tilde \tau$'s are defined as in Eq.~\ref{equ:pweight3}.)

Next, consider the \textbf{Green function}, $G(\tau' - \tau)=-Z^{-1}\sum_\mathcal{C}w(\mathcal{C}, c(\tau') \overline{c}(\tau))$. Here, the assumption that $w(\mathcal{C}, c(\tau') \overline{c}(\tau))$ vanishes when $w(\mathcal{C})$ vanishes is problematic due to the Pauli exclusion principle. Fortunately, this can be remedied: Instead inserting two operators in order to recover the Green's function weight, we instead remove the hybridisation lines from two operators. The resulting estimator reads
\begin{equation}
\label{equ:PGreenEstimator}
G_{ij}(i\nu_n) = -\frac{\partial \ln Z}{\partial \Delta_{ji}(i\nu_n)}= -\beta^{-1} \biggl \langle   \sum_{r, s=1}^{|\mathcal C |}e^{i\nu_n(\tau_s' - \tau_r)}\delta_{ii'_s}\delta_{ji_r} (M^{-1})_{sr} \biggr \rangle_{w(\mathcal C)/Z}
\end{equation}
where $M_{rs}=\Delta_{i_ri_s'}(\tau_r - \tau_s')$. This approach solves our problem. However, it introduces limitations. Namely, only entries $G_{ij}$ of the Green function for which $\Delta_{ji}$ are non-zero can be calculated with this estimator, and this estimator may fail if the bath has very few orbitals, as discussed in \ref{app:EDimpurities}. DMFT calculations are in general not affected by these limitations.

The most important observable for DMFT calculations is the \textbf{self-energy}, which can be calculated with the Green function and the Dyson equation, $\Sigma = \mathcal{G}^{-1} - G^{-1}$. This is numerically unfavorable, as it amplifies the statistical noise in the Green function at higher frequencies: $\delta \Sigma = G^{-2}\delta G \propto \nu_n^2 \delta G$. A better method calculates the following correlation function 
\begin{equation} 
\label{equ:ImprCorrFunc}
H_{ij}(\tau) = - \langle T_\tau  [\hat c_i, \hat U](\tau) \hat{c}_j^\dagger \rangle - \int_0^\beta \sum_{IJ}q_{iI}D_{IJ}(\tau - \tau')\langle T_\tau \hat c_i(\tau) \hat Q_J(\tau') \hat{c}_j^\dagger \rangle d\tau'
\end{equation}
and obtains the self-energy from $\Sigma = HG^{-1}$.\cite{Hafermann2012,Gunacker2016} Here $\hat{U}$ is the interacting part of $\hat{H}_\text{loc}$ in Eq.~\ref{equ:ImpurityHamiltonian}. This reduces the noise at the higher frequencies, as $\delta \Sigma = G^{-2}H\delta G + G^{-1}\delta H$, which roughly goes as $\nu_n$ times the noise in $G$ and $H$.

The estimator for Eq.~\ref{equ:ImprCorrFunc}, that is, the \textbf{improved estimator for the self-energy}, has a contribution from the static and the retarded part of the interaction. It reads
\begin{equation}
\begin{split}
H_{ij}(i\nu_n) = -\beta^{-1} \biggl \langle   \sum_{r,s=1}^{|\mathcal C |}e^{i\nu_n(\tau_s' - \tau_r)}\delta_{ii'_s}\delta_{ji_r} (M^{-1})_{sr} \times (h^{\text{st}}_{s} + h^{\text{ret}}_{s})\biggr \rangle_{w(\mathcal C)/Z}.
\end{split}
\end{equation}
The contribution from the static part, $h^{\text{st}}_{s}$, is
\begin{equation}
\label{equ:PImprovedEstimatorStatic}
h^{\text{st}}_{s} = \frac{w_{\text{loc},s}^U(\mathcal C)}{w_\text{loc}(\mathcal C)},
\end{equation}
where $w_{\text{loc},s}^U(\mathcal C)$ is obtained from $w_\text{loc}(\mathcal C)$ by replacing $\hat{c}_{i_s'}(\tau_s')$ with the Bulla operator $[\hat c_{i_s'}, \hat U](\tau_s')$. See, for example, Ref. \cite{Hafermann2012,Gunacker2016}. The contribution from the retarded part, $h^{\text{ret}}_{s}$ is 
\begin{equation}
h^{\text{ret}}_{s} =  \sum_{IJ} q_{i_s'I} \biggl (D_{IJ}(i\omega_n = 0) \frac{w_\text{loc}(\mathcal C, Q_J)}{w_\text{loc}(\mathcal C)} - \sum_{t=1}^{2|\mathcal C|} \tilde q_{tJ}L_{I J}(\tau_s' - \tilde{\tau}_t) \bigg ),
\end{equation}
where $L_{IJ}(\tau)$ is the first primitive of $D_{IJ}(\tau)$, and the $\tilde q$'s and $\tilde \tau$'s are defined as in Eq.~\ref{equ:pweight3}. While it is not clear \textit{a priori} that $w_{\text{loc},s}^U(\mathcal C)$ vanishes whenever $w_\text{loc}(\mathcal C)$ vanishes, we have not encountered any problems where this assumption is violated. Notice in this respect that, at least from the point of view of their symmetries, $w_\text{loc}(\mathcal C)$ and $w_{\text{loc},s}^U(\mathcal C)$ have the same zero's. This is because $[\hat{c}_i, \hat{U}]$ transforms as $\hat{c}_i$ under the symmetries of $\hat{H}_\text{loc}$, since we can assume that $\hat{U}$ transforms as the identity.

Until now, we assumed that the weights in the partition function expansion are positive in order to interpret $w(\mathcal C)/Z$ as a probability distribution. If $w(\mathcal C)$ becomes negative or complex, which is generally the case since we are dealing with fermions, we rewrite Eq.~\ref{equ:PartitionOExpansion} as
\begin{equation}
\label{equ:PSignRewrite}
\langle \mathcal O \rangle = \frac{|Z|}{Z}\sum_\mathcal{C} \frac{|w(\mathcal C)|}{|Z|} \frac{w(\mathcal C)}{|w(\mathcal C)|}o(\mathcal C, \mathcal O) =\frac{|Z|}{Z}\biggl \langle \frac{w(\mathcal C)}{|w(\mathcal C)|} o(\mathcal C, \mathcal O) \biggr \rangle_{|w(\mathcal C)|/|Z|}
\end{equation}
and sample the ``bosonic'' partition function $|Z|=\sum_\mathcal{C} |w(\mathcal C)|$. The estimators we discussed above get a phase factor $w(\mathcal C)/|w(\mathcal C)|$, and we need to estimate the ratio $Z/|Z|$, which is also called the ``sign'' of the Monte-Carlo simulation. The sign is not a scalar observable in the proper sense and depends for example on the one particle basis \cite{SemonSubleadingCorr,ShinaokaSign}. 

The estimator of the \textbf{sign} is  
\begin{equation}
\label{equ:Sign}
\frac{Z}{|Z|} = \biggr \langle \frac{w(\mathcal C)}{|w(\mathcal C)|} \biggr \rangle_{|w(\mathcal C)|/|Z|}.
\end{equation}
Notice that while $w(\mathcal C)/|w(\mathcal C)|$ can be complex, the sign is always real because the partition function $Z$ is real.
In real materials, it can be shown that the sign becomes exponentially small upon lowering the temperature, whereas the variance of the random variable $w(\mathcal C)/|w(\mathcal C)|$ gets closer and closer to one. The exponential blow up of the statistical noise in Eq.~\ref{equ:PSignRewrite} that this entails is called the ``sign problem''. That is, each measurement provides $Z/|Z|$ information, and so one must sample an additional $(Z/|Z|)^{-1}$ configurations in order to receive a good estimate of any observable. A sign below 0.1 is generally considered unacceptable, as it indicates at least an order of magnitude more computational resources must be dedicated to the CTQMC.
   
The approach of removing hybridisation lines as in Eq.~\ref{equ:PGreenEstimator} can also be used to obtain estimators for higher order correlation functions, for example for the two particle Green function $\langle c_i \overline{c}_j c_k \overline{c}_l\rangle = -Z^{-1}\partial^2 Z /\partial \Delta_{ji}\partial \Delta_{lk}$. Similar to the one particle Green function, the limitation is that both $\Delta_{ji}$ and $\Delta_{kl}$ are non-zero. As a consequence, not all entries of the two particle Green function can be accessed. Consider, for example, an impurity model with a diagonal Weiss Field and a Kanamori interaction: entries with $i\ne j \ne k \ne l$ can be finite due to spin-flip processes, while only entries with $i=j$ and $k=l$ (or $i=l$ and $j=k$) can be accessed since the hybridisation is diagonal. To overcome this limitation, one must sample the configuration space of not only the partition function, but also the observable. This is called worm sampling. Let us discuss.
       
\subsection{Worm-CTQMC}\label{sec:Worms}

Let us briefly outline the theory behind worm sampling using the one particle Green function as our observable. We begin with the observable expansion, Eq.~\ref{equ:Oexpansion}, for $\mathcal{O}=c(\tau)\overline{c}(\tau')$. Let us also include the Fourier transform, so that we sample in the Matsubara basis.  Sampling this expansion yields an estimate of the ``Green function''   
\begin{equation}
\label{equ:GreenWormExpansion}
\begin{split}
\tilde{G}(i\nu_n) &= -(\beta |Z_G|)^{-1}\sum_{\mathcal C, \tau\tau'}w(\mathcal C, c(\tau)\overline{c}(\tau')) e^{i\nu_n (\tau - \tau')} \\
&=  -\beta^{-1} \biggl \langle  \frac{w(\mathcal C, c(\tau)\overline{c}(\tau'))}{|w(\mathcal C, c(\tau)\overline{c}(\tau'))|} e^{i\nu_n (\tau - \tau')} \biggr \rangle_{|w(\mathcal C, c(\tau)\overline{c}(\tau'))|/|Z_G|}.
\end{split}
\end{equation}
This function is not normalized by the partition function, $Z$, like the true Green's function. Instead, it is normalized to $|Z_G|=\sum_\mathcal{C,\tau\tau'} |w(\mathcal C, c(\tau) \overline c(\tau'))|$. To get the real Green's function, then, we need an estimate of $|Z_G|/Z$. To this end, we consider an extended sampling space which consists of both the Green function space $\mathcal{S}_G=\{(\mathcal C, \tau\tau' )\}$ and the partition function space $\mathcal{S}_Z=\{ \mathcal C \}$, with probability distribution $|w(\mathcal C, c(\tau)\overline{c}(\tau'))|/(|Z_G| + |Z|)$ and $|w(\mathcal C)|/(|Z_G| + |Z|)$, respectively. The number of samples $N_G$ and $N_Z$ taken in each space when sampling $\mathcal S = \mathcal S_G \cup \mathcal S_Z$ is then proportional to $|Z_G|$ and $|Z|$, so that
\begin{equation}
G(i\nu_n) = \frac{|Z|}{Z}\times \frac{|Z_G|}{|Z|}\times \tilde{G}(i\nu_n) \approx \frac{|Z|}{Z}\times \frac{N_G}{N_Z}\times \tilde{G}(i\nu_n).
\end{equation}
Notice that this extended sampling simultaneously yields estimates of the ``Green function'' in Eq.~\ref{equ:GreenWormExpansion} (when in $\mathcal S_G$), the sign $Z/|Z|$, as well as any other observables one can sample in partition space (when in $\mathcal S_Z$), and the relative volume $|Z_G|/|Z|$ of $\mathcal S_G$ and $\mathcal S_Z$.  The operators $c(\tau) \overline c(\tau')$ are called ``worm'' operators, as they are said to worm between the partition and observable spaces. Thus, we arrive at the name Worm-CTQMC. 

\subsubsection{Correlation Functions}\label{sec:WormCorrelationFunctions}

In this section, we list the correlation functions for which worm sampling is implemented. These are 
\begin{align}
 G^{(1)}_{ij}(\tau_{12}) &=-\langle T_\tau \hat{c}_i(\tau_1)\hat{c}^\dagger_j(\tau_2) \rangle &(*) \label{equ:CorrFuncStart} \\
 G^{(2)}_{ijkl}(\tau_{12},\tau_{34}, \tau_{23}) &=-\langle T_\tau \hat{c}_i(\tau_1)\hat{c}_j^\dagger (\tau_2) \hat{c}_k(\tau_3)\hat{c}_l^\dagger(\tau_4) \rangle &(*) \\
 G^{(2,ph)}_{ijkl}(\tau_{12},\tau_{23})&= -\langle  T_\tau \hat{c}_i(\tau_1)\hat{c}_j^\dagger(\tau_2) \hat{n}_{kl}^{(ph)}(\tau_3)\rangle &(*)\label{equ:twothreeGreenPH} \\
 G^{(2,pp)}_{ijkl}(\tau_{12},\tau_{23})&= -\langle  T_\tau \hat{c}_i(\tau_1)\hat{c}_j(\tau_2) \hat{n}_{kl}^{(pp)}(\tau_3) \rangle &(*) \label{equ:twothreeGreenPP} \\
 G^{(2,ph)}_{ijkl}(\tau_{12}) &= -\langle T_\tau \hat{n}_{ij}^{(ph)} (\tau_1) \hat{n}_{lk}^{(ph)}(\tau_2)  \rangle & \\
 G^{(2,pp)}_{ijkl}(\tau_{12})&= -\langle T_\tau \hat{n}_{ij}^{(hh)}(\tau_1)  \hat{n}_{kl}^{(pp)}(\tau_2) \rangle & \label{equ:CorrFuncEnd} 
\end{align}
where $\hat{n}_{ij}^{(ph)}=\hat{c}_i^\dagger \hat{c}_j$, $\hat{n}_{ij}^{(pp)}=\hat{c}_i^\dagger \hat{c}_j^\dagger$, $\hat{n}_{ij}^{(hh)}=\hat{c}_i\hat{c}_j$ and $\tau_{12}=\tau_1 - \tau_2$. For the Green functions marked with an asterisk, improved estimators are implemented. The higher-order correlation functions are obtained by replacing the first operator $\hat{c}_i(\tau_1)$ in the respective Green function with the Bulla operator $[\hat{c}_i, \hat{U}](\tau_1)$ and adding the contribution coming from
the retarded interaction, that is,
\begin{equation}
\label{equ:GenHigherOrderCorrelator}
H(\dots) = -\langle T_\tau [\hat{c}_i, \hat{U}](\tau_1)\cdots \rangle - \int_0^\beta \sum_{IJ}q_{iI}D_{IJ}(\tau_1 - \tau')\langle T_\tau \hat c_i(\tau_1) \hat Q_J(\tau') \cdots \rangle d\tau',
\end{equation}
where the $\cdots$ denotes the operators other than $\hat{c}_i(\tau_1)$ in the Green function. For an example, see  Eq.~\ref{equ:ImprCorrFunc}, which gives the results for the one-particle Green function.

\subsubsection{Sampling}
As outlined above for the one-particle Green function, Worm-CTQMC extends the configuration space by adding additional, worm spaces to the partition space. That is, we must sample an additional worm space for each correlation function that needs to be sampled.  In general, the additional worm spaces may be much larger or smaller than each other or than partition space. In this case, the Worm-CTQMC will spend the vast majority of its time in the largest configuration spaces. To compensate, the volume of each space is normalized by an additional term,  $\eta_{\mathcal{X}}$. Therefore, we sample the union of normalized spaces
\begin{equation}
\label{equ:WormSpaces}
\mathcal S = \bigcup_{\mathcal{X}} \eta_{\mathcal{X}} \mathcal S_{\mathcal{X}}.
\end{equation}
In our case, the union goes over the Green's functions listed in Eqs.~\ref{equ:CorrFuncStart} to \ref{equ:CorrFuncEnd}, and the corresponding higher-order correlators defined in Eq.~\ref{equ:GenHigherOrderCorrelator}, and the partition function $\mathcal X=Z$. The probability distribution on $\eta_{\mathcal X} \mathcal S_{\mathcal X}$ is proportional to $\eta_{\mathcal X} \times |w(\mathcal C, \mathcal O)|$, where $\eta_{\mathcal X}$ is discussed in more detail below, and $\mathcal O$ is given as follows:
\begin{itemize}[itemindent=3em]
\item[($\mathcal X = G$):] For the Green functions, $\mathcal O$ represents the operators which define the function, e.g., $\mathcal O = \hat{c}_i(\tau_1)\hat{c}_j^\dagger(\tau_2)  \hat{n}_{kl}^{(ph)}(\tau_3)$ for $G^{(2,ph)}(\tau_{12},\tau_{23})$ in Eq.~\ref{equ:twothreeGreenPH}. 
\item[($\mathcal X = H$):] For the improved estimators, $\mathcal O$ represents the operators which define the static part of Eq.~\ref{equ:GenHigherOrderCorrelator}, e.g., $\mathcal O = [\hat{c}_i, \hat{U}](\tau_1)\hat{c}_j(\tau_2) \hat{n}_{kl}^{(pp)}(\tau_3)$ for $H^{(2,pp)}(\tau_{12},\tau_{23})$. The dynamic part of the improved estimator is measured in the associated Green's function space. 
\item[($\mathcal X = Z$):] For the partition function $\eta_{\mathcal X}=1$ (by convention) and the probability distribution is proportional to $|w(\mathcal C)|$.
\end{itemize} 

When sampling the space $\mathcal S$ with a Markov-Chain using the Metropolis-Hasting algorithm, a new configuration $(\mathcal C', \mathcal O')$ is proposed according to the trial probability $q(\mathcal C', \mathcal O' | \mathcal C, \mathcal O)$, and accepted with probability
\begin{equation}
p=\text{min}\biggl (\frac{q(\mathcal C', \mathcal O' | \mathcal C, \mathcal O)}{q(\mathcal C, \mathcal O|\mathcal C', \mathcal O')}\times \frac{\eta_{\mathcal X'}}{\eta_{\mathcal X}} \times \biggl | \frac{w(\mathcal C', \mathcal O')}{w(\mathcal C, \mathcal O)} \biggr |,\,\, 1 \biggr).\label{equ:worm metropolis hastings}
\end{equation}
The trial configuration $(\mathcal C', \mathcal O')$ can lie in the same space as the present configuration ($\mathcal  X=\mathcal X'$) or not ($\mathcal X'\ne \mathcal X$). The present CTQMC solver implements the following moves:
\begin{itemize}[itemindent=3em]
\item[($\mathcal X\ne \mathcal X'$):] Insert or remove the worm to connect the worm space with the partition space, that is, $(\mathcal C, \mathcal O)\leftrightarrow (\mathcal C, \varnothing)$. The flavors and imaginary times of the inserted worm operators are drawn uniformly.
\item[($\mathcal X\ne \mathcal  X'$):] Insert or remove a $\hat{c}(\tau)\hat{c}^\dagger(\tau')$, $[\hat{c}, \hat{U}](\tau)\hat{c}^\dagger(\tau')$ or $\hat{n}^{(ph)}(\tau)$ from the worm to connect different worm spaces (if compatible), e.g., $(\mathcal C, \mathcal O)\leftrightarrow (\mathcal C,  \mathcal O + \hat{c}(\tau)\hat{c}^\dagger(\tau'))$ for $\mathcal O \in G^{(1)}$ or $H^{(1)}$. The flavors and imaginary times of the inserted operators are drawn uniformly.
\item[($\mathcal X = \mathcal X'$):] Swap the imaginary time of an $\hat{c}(\tau)$ or $\hat{c}^\dagger(\tau)$ operator in the worm and a partition operator of the same flavor, e.g., $(\hat{c}(\tau)\cdots, \hat{c}(\tau')\cdots)$ $\leftrightarrow$ $(\hat{c}(\tau')\cdots, \hat{c}(\tau)\cdots)$.  Also implemented is a related move for the $\hat{n}^{(ph)}(\tau)$,  $\hat{n}^{(pp)}(\tau)$ and $\hat{n}^{(hh)}(\tau)$ operators in a worm. See\ref{sec:CTQMCMoveOptimizations} for details.
\item[($\mathcal X = \mathcal X'$):] Partition space configuration moves in all spaces, that is, $(\mathcal C, \mathcal O)\leftrightarrow (\mathcal C', \mathcal O)$. 
\end{itemize}
For the $\mathcal X \ne \mathcal X'$ moves listed above, the contribution $w_\text{hyb}$ from the hybridisation to the weight $w$ in the Metropolis-Hasting acceptance ratio cancels and is therefore not calculated.

\begin{figure}
\centering
\includegraphics[scale=1]{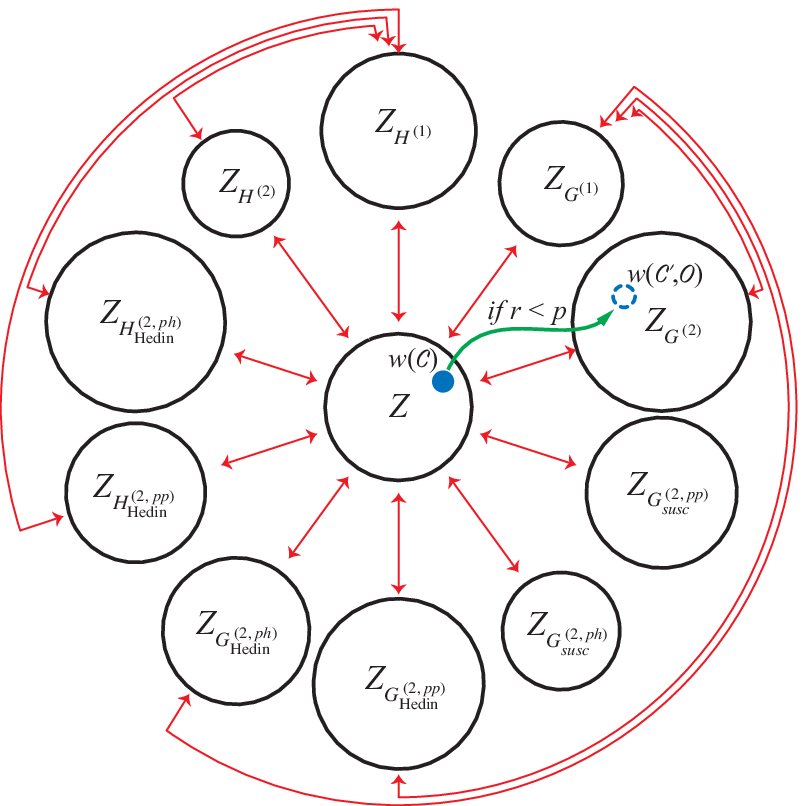}
\caption{The various configuration spaces of \ctqmcname{} and their connections. The Markov chain moves between or within a configuration space by comparing the relative weight of the current and proposed configuration while accounting for the detailed balance and the renormalized volume of the starting and ending configuration spaces. (See Eq. \ref{equ:worm metropolis hastings} and Eq. \ref{equ:etas}. The proposed move is accepted if a uniform random number $r\in [0,1)<p$.)
 }\label{fig:ConfigSpaces}
\end{figure}

Figure \ref{fig:ConfigSpaces} illustrates the configuration spaces sampled by \ctqmcname{}. The spaces which \ctqmcname{} can worm between are connected by red lines. Also shown is a Markov chain in some configuration. $\mathcal{C}$, proposing a move which worms between the partition and two-particle Green's function space. (That is, a move which inserts the four worm operators $c_i\overline{c}_jc_k\overline{c}_l$ at random points on the imaginary time axis.)
Also shown is a depiction of the disparity in volume between various configuration spaces. In reality, the  volume of one configuration will often be orders of magnitude larger than the others, which is why we must scale the relative volume of each space.
The scaling factors $\eta_{\mathcal X}$ in Eq.~\ref{equ:WormSpaces} are chosen such that the Markov-Chain spends approximately the same amount of time in each space, that is, 
\begin{equation}
\eta_{\mathcal X} \approx \frac{|Z| }{|Z_{\mathcal X}|},\label{equ:etas}
\end{equation}
where $|Z_{\mathcal X}| = \sum_{\mathcal C, \mathcal O\in \mathcal X} |w(\mathcal C, \mathcal O)|$. This is achieved with a Wang-Landau algorithm \cite{Wang2001a,Wang2001b,Shinoaka2017b}.

\subsubsection{Estimators}\label{sec:worm observables}
In this section, the estimators for the correlation functions listed in Sec.~\ref{sec:WormCorrelationFunctions} are discussed. To this end, we start with a detailed discussion of the estimator for the one-particle Green function and the associated improved estimator. 

The estimator for the one-particle Green function reads  
\begin{equation}
G_{ij}(i\nu_n) =  -\beta^{-1} \mathcal{N}_G \langle  w/|w|\times \delta_{ii'}\delta_{jj'}e^{i\nu_n (\tau - \tau')} \rangle_{|w(\mathcal C,G )|/|Z_G|}.
\end{equation}
The subscript denotes that this measurement is averaged across all measurements taken in in the Green's function configuration space. Therefore, the normalization factor is given by
\begin{equation}
\mathcal{N}_G = \frac{|Z_G|}{Z} \approx \frac{|Z|}{Z}\times \frac{N_G}{\eta_G N_Z}.
\end{equation}
Here $N_Z$ and $N_G$ are the number of samples taken in the partition space and the Green function space, respectively. For the higher-order correlation function $H=H^{\text{st}}+H^{\text{ret}}$ in Eq.~\ref{equ:GenHigherOrderCorrelator}, the static part is estimated from the $\mathcal S_{H}$ space,
\begin{equation}
H^{\text{st}}_{ij}(i\nu_n) =  -\beta^{-1} \mathcal{N}_H \langle  w/|w|\times \delta_{ii'}\delta_{jj'}e^{i\nu_n (\tau - \tau')} \rangle_{|w(\mathcal C, H)|/|Z_H|},
\end{equation}
while the retarded part is estimated from the \textit{Green function space} $\mathcal S_{G}$,
\begin{equation}
H^{\text{ret}}_{ij}(i\nu_n) =  -\beta^{-1} \mathcal{N}_G \langle  w/|w|\times \delta_{ii'}\delta_{jj'}e^{i\nu_n (\tau - \tau')} \times h \rangle_{|w(\mathcal C, G)|/|Z_G|},
\end{equation}
where
\begin{equation}
h =  \sum_{IJ} q_{i'I} \biggl (D_{IJ}(i\omega_n = 0) \frac{w_\text{loc}(\mathcal C, \hat{c}_{i'}(\tau)\hat{c}_{j'}^\dagger(\tau') \hat{Q}_J)}{w_\text{loc}(\mathcal C, \hat{c}_{i'}(\tau)\hat{c}_{j'}^\dagger(\tau'))} - \sum_{t=1}^{2|\mathcal C| + 2} \tilde q_{tJ}L_{I J}(\tau - \tilde{\tau}_t) \bigg ).
\end{equation}
Here $\tilde \tau_t$ and $\tilde{q}_{tI}$ are as defined in Eq.~\ref{equ:pweight3} for $t=1,\dots,2|\mathcal C|$. For $t=2|\mathcal C| + 1$ we set $\tilde \tau_t =\tau$ and $\tilde q_{tI}=q_{i'I}$ (which comes from $\hat{c}_{i'}$), and for $t=2|\mathcal C| + 2$ we set $\tilde \tau_t =\tau'$ and $\tilde q_{tI}=-q_{j'I}$ (which comes from $\hat{c}_{j'}^\dagger$).

The estimators for the other correlation functions are obtained analogously, and we list  just the estimators for the other Green functions  
\begin{align}
G^{(2)}(i\nu_n,i\nu_{n'},i\omega_m) &= -\mathcal{N}_{G^{(2)}} \langle  w/|w|\times e^{i\nu_{n}\tau_{12} + i\nu_{n'}\tau_{34} + i\omega_m\tau_{23}}\rangle_{G^{(2)}} \\
G^{(2,l)}(i\nu_n,i\omega_m) &=-\mathcal{N}_{G^{(2,l)}(2)}\langle  w/|w|\times e^{i\nu_n \tau_{12}+i\omega_m \tau_{23}}\rangle_{G^{(2,l)}(2)}\\
G^{(2,l)}(i\omega_m) &= -\mathcal{N}_{G^{(2,l)}(1)} \langle w/|w|\times e^{i\omega_m \tau_{12}}\rangle_{G^{(2,l)}(1)}
\end{align}
where $l=ph,pp$ and the flavor indices are omitted.
We store the estimators for all worm-based observables in either in this Fourier basis \cite{Gunacker2015,Gunacker2016,Kaufmann2017} or mixed Legendre-Fourier basis \cite{Boehnke2011}. For the technical details of the mixed representation, we refer the reader to Ref.~\citen{Boehnke2011}. 

\subsubsection{Susceptibilities}
\ctqmcname{} converts all worm-based estimators and improved estimators into susceptibilities, $G_{ijkl},H_{ijkl}\rightarrow\chi_{ijkl}$, during a post-processing step. To do this, the disconnected part of the susceptibility is computed as
\begin{align}
\chi_{\mathrm{disc},ijkl}^{(ph)}(i\omega_m) &= n_i n_k \delta_{ij} \delta_{kl}\delta_{0m}\\
\chi_{\mathrm{disc},ijkl}^{(pp)}(i\omega_m) &= 0 \\
\chi_{\mathrm{disc},ijkl}^{(ph)} (i\nu_n,i\omega_m)&=  G_{ii}(\nu_n)[n_l\delta_{ij}\delta_{kl}\delta_{0m} - G_{ll}(i\nu_n-i\omega_m)\delta_{ik}\delta_{jl}]\\
\chi_{\mathrm{disc},ijkl}^{(pp)}(i\nu_n,i\omega_m) &= G_{ii}(i\nu_n)G_{kk}(i\omega_m-i\nu_n)(\delta_{ik}\delta_{jl}-\delta_{il}\delta_{jk}) \\
\chi_{\mathrm{disc},ijkl}(i\nu_n,i\nu_{n'},i\omega_m) &=  \delta_{ij}\delta_{kl}\delta_{0m} G_{ii}(i\nu_n)G_{kk}(i\nu_{n'}) \notag\\
& - \delta_{il}\delta_{jk}\delta_{nn'} G_{ii}(i\nu_n)G_{kk}(i\nu_n-i\omega_m).
\end{align}
Note that these two- and three-point definitions differ from the typical expressions, e.g., those written in Ref. \cite{Kaufmann2017}. This discrepancy arises because we write our observables, Eqs. \ref{equ:CorrFuncStart} to \ref{equ:CorrFuncEnd}, using a different operator ordering. Our ordering allows us to define the operators in terms of bilinears, which simplifies the implementation.  Note that the occupations $n_i$ are measured in the partition space, and the Green's functions can be measured in either partition space or in the Green's function worm space. \ctqmcname{} will use the worm-space measurements if they are available and the partition-space measurements if they are not.

Once the disconnected parts are computed, the susceptibility is computed from the estimator or improved estimator as\cite{Gunacker2016}
\begin{align}
\chi_{ijkl} &= G_{ijkl}-\chi_{\mathrm{disc},ijkl} \\
\chi_{ijkl}&=\sum_m \frac{G_{mi}(i\nu_n)\Sigma_{mi}(i\nu_n)\chi_{\mathrm{disc},ijkl} - G_{mi}(i\nu_n)H_{ijkl}}{\delta_{im}+G_{mi}\Sigma_{mi}(i\nu_n)},
\end{align}
respectively, where $(i\nu_n)$ is the first fermionic frequency of the three- or four-point susceptibility $\chi_{ijkl}$. (We do not compute improved estimators for the two-point susceptibility.)

With the theory established, let us discuss our implementation.

%% file: Src/Implementation.tex
In this section, we present the various techniques implemented in \ctqmcname{} in order to reduce or overcome the computational burden. Here we will discuss the parallelism of \ctqmcname{} and the GPU acceleration of the CTQMC algorithm. Details regarding the optimization of the Monte-Carlo moves are available in \ref{sec:CTQMCMoveOptimizations}, and the application of symmetries which help with the time-to-solution for two-particle correlators are given in \ref{app:susceptibilities}. 
Before discussing these topics, however, it is useful to outline the phases of the CTQMC simulation. 

\subsection{Structure}\label{structure}

\begin{figure}
\centering
\includegraphics[scale=1]{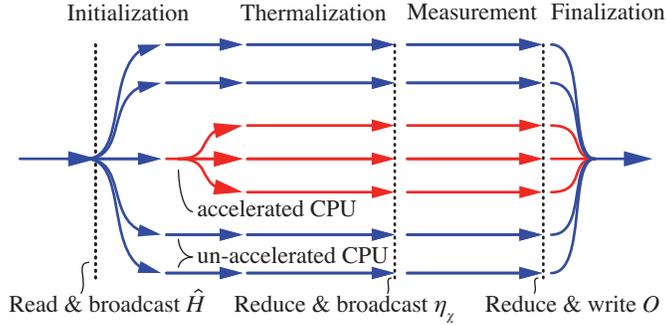}
\caption{ A simple illustration of the parallelism and structure of CTQMC. Each blue path represents the work of a single un-accelerated CPU, i.e., it represents a simulation. The red paths together  represent the work of a single accelerated CPU (which handles, in this illustration, three simulations). The simulations proceed through four phases: initialization, thermalization, measurement, and finalization. Three points of CPU-CPU communication occur in these phases, as indicated by a dotted lines: (1) The control parameters and impurity problem are broadcast to each CPU; (2) the scaling factors, $\eta_\chi$, are averaged across all simulations and then broadcast back to all ranks; and (3) the results are gathered from all simulations. }
\label{fig:structure}
\end{figure}

In \ctqmcname{}, the CTQMC simulation is divided into four major phases: (1) Initialization, (2) thermalization, (3) measurement, and (4) finalization. These phases are listed in Fig. \ref{fig:structure}, which depicts the general parallelism scheme. 

(1) During initialization, \ctqmcname{} reads to the input files, distributes the impurity problem and control parameters to all CPUs, initializes the GPU, and pre-computes those quantities which remain static throughout a CTQMC simulation (e.g., it decomposes the Hilbert space into invariant spaces, computes the eigenvalues and eigenvectors of these spaces, and generates the operator matrices associated with these eigenstates.) 

(2) During thermalization, the simulations (Markov chains) are run for a user specified number of minutes or steps, and the Wang Landau algorithm is used in order to determine the relative size of each configuration space and compute the resulting set of $\eta_{\chi}$. The goal in this phase, aside from gathering the set of $\eta_{\chi}$, is to reach a physical region of the configuration space.

(3) During measurement, the simulations are run for a user specified number of minutes or steps, as in the thermalization phase. Instead of computing $\eta_{\chi}$, however, all of the observables are sampled. The goal in this phase is for the estimators associated with these oberservables to converge. 

(4) During finalization, the results from each simulation are gathered, and the error is estimated (using the set of results from all Markov chains simulated or the previous set of results). Then, the results are written to a file. Additionally, the final configuration of each simulation is written to file. This allows subsequent runs of \ctqmcname{} to eliminate or at least substantially reduce the length of the thermalization phase. 

Now, let us discuss the parallelism and how it relates to these phases.

\subsection{Parallelism}

One of the major reasons to use CTQMC is its nearly ideal weak scaling. That is, one can simulate many independent Markov chains and then collect the results from each simulation. Very little or no communication between these simulations is required, as shown in Fig. \ref{fig:structure}. Therefore, a CTQMC code can simulate one Markov chain per CPU without incurring a noticeable performance loss. \ctqmcname{} uses the message passing interface (MPI) to initialize and distribute these Markov chains, collect the results, and compute the average $\eta_{\mathcal{O}}$ after thermalization. A small amount of serial code is required before or after these communication events, e.g., to compute the statistics of the results (across the many Markov chains simulated), conduct I/O, and decompose the Hilbert space of the local Hamiltonian into invariant subspaces, as illustrated in Fig. \ref{fig:structure}. 

However, one should note that the solution is being worked on only during the measurement phase; therefore, the remaining phases essentially do not scale with the number of workers. Still, a substantial thermalization phase is only required for the first CTQMC run, and the initialization and finalization phases tend to be very quick. (The decomposition of the Hilbert space and creation of the operator matrices can take a lot of computation in hardest, low-symmetry $f$-shell systems. However, this computation can be parallelized across a few nodes.)

\begin{figure}
\centering
\includegraphics[scale=1]{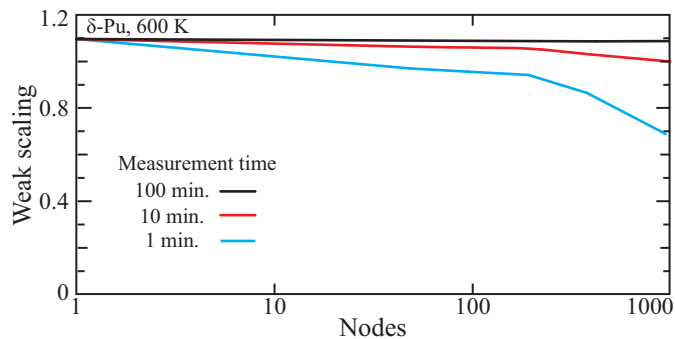}
\caption{ The weak scaling of \ctqmcname{} with the number of nodes for variations in the length of the measurement phase. The lack of communication during this phase leads to nearly ideal scaling, provided the measurement is at least 10 minutes long. Tests were run on the supercomputer Summit at the Oak Ridge National Laboratory. These results assume that the simulations are already thermalized.
}
\label{fig:scaling}
\end{figure}

The structure of this parallelism is simple and effective. As no CPU-to-CPU communication occurs during the computationally demanding measurement or thermalisation phases of the CTQMC, \ctqmcname{} scales extremely well with the number of CPUs. Indeed, if the measurement time is long, \ctqmcname{} scales ideally, as shown in Fig. \ref{fig:scaling}(a):  Provided the user does not require measurement across thousands of CPUs within 10 minutes, the scaling remains above 0.90 of the ideal. We reiterate that this feature of CTQMC is one of the reasons it has risen to prominence, and is not unique to \ctqmcname{}. 

Now, let us discuss the acceleration of the CTQMC algorithm by a GPU. 

\subsection{GPU acceleration of the local trace}\label{sec:trace}

When one seeks to use GPUs to accelerate an algorithm, it is critical to identify the computational bottleneck. Then, one must ensure that this bottleneck involves the manipulation of large matrices. In impurity solvers,  one must  deal with a Hilbert space that scales as $d=2^n$, where $n$ is the number of orbitals.  In completely un-optimized CT-HYB, computing the local trace involves multiplying $2k$ matrices of rank $2^n$, where $k$ is the expansion order of the Markov chain, i.e., $2k=|C|+|O|$. In an $f$-shell problem, $n=14$, the Hilbert space has a dimension of 16,384, and solving the impurity problem becomes impossible. (In comparison, $d$-shell impurities, $n=10$, have a Hilbert space of rank 1024.)  While many optimizations have been adopted by current CTQMC solvers which allow us to solve the impurity problem \cite{haule2007quantum, gull2008continuous, yee2012towards, semon2014lazy, seth2016triqs, shinaoka2014hybridization}, the local trace remains the bottleneck due to this underlying scaling of the Hilbert space. This is particularly true in $d$-shell and $f$-shell problems at moderate-to-high temperatures.  Fortunately, this scaling also implies that the algorithm may be accelerated by using GPUs to compute the local trace.

Before discussing in detail how we use GPUs to accelerate the trace computation, it will be useful to outline more concretely how this computation is handled in \ctqmcname{}. \ctqmcname{} employs the lazy skip-list\cite{semon2014lazy} algorithm to optimize the trace computation.  Here we only provide a brief overview of the algorithm; for a detailed description of the algorithms, we refer the reader to Ref. \cite{semon2014lazy}. 

\subsubsection{Computing the local impurity trace}

In a naive implementation of CTQMC, the local impurity trace in Eq. \ref{equ:pweight1}, is evaluated as follows. First, one represents the set of operators in the many-bodied Hilbert space, $\mathcal{H}$, via the basis $(F_i)_{nm}=\langle m | \hat{c}_i | n  \rangle$, where $|n\rangle$ and $|m\rangle$ are states in the Hilbert space. Then, the trace can be represented as 
\begin{align}
w_{\mathrm{loc}}(\mathcal{C}) = \mathrm{Tr} P_{\beta-\tau_k}F^\dagger_{i_k}P_{\tau_k-\tau_k'}F_{i_k'} \cdots F_{i_1} P_{\tau_1-\tau_1'}F^\dagger_{i_1}P_{\tau_1'},
\end{align}
where $P_\tau=\exp(-\tau H_{\mathrm{loc}})$ is the (diagonal) projector matrix and we have already applied the time ordering operator. In this un-optimized implementation, the computational burden is $O[k (2^n)^3]$, where $n$ is the number of flavors in the impurity problem. (That is, we must compute order $k$ matrix products for matrices of rank $2^n$.) This calculation is essentially impossible for an $f$-shell problem where $n=14$. 

Fortunately, the local Hamiltonian contains Abelian symmetries associated with some quantum numbers, e.g., the particle number, $S_z$, $S^2$, $J_z$, $J^2$, etc. This allows us to decompose the Hilbert space into a series of sectors, 
\begin{align}
\mathcal{H} = \bigoplus_{q}\mathcal{H}(q),
\end{align}
where $q$ enumerates the sectors of the Hilbert space, each with its own unique set of quantum numbers. Moreover, it allows us to define a new basis for the creation and annihilation operators, $F_{\alpha}(q_i)$, which \textit{uniquely} maps one sector onto another sector, $q_i\rightarrow q_{i+1}$. (We are limited to Abelian symmetries, because we require the uniqueness of this mapping.) 
Therefore, we can define the operator matrices $[F_\alpha(q_i)]_{nm}=\langle m(q_{i+1}) | \hat{c}_\alpha | n(q_{i})  \rangle$, where $|m(q_{i+1})\rangle$ and $|n(q_{i})\rangle$ are many-bodied states in the Hilbert subspaces (or sectors) $q_{i+1}$ and $q_i$, and $\alpha$ specifies both the operator flavor and type (annihilation or creation). In this basis, we can rewrite the local trace as a sum over a set of initial sectors, $q_0$. That is,
\begin{align}
w_{\mathrm{loc}}(\mathcal{C}) = \sum_{q_0}\mathrm{Tr} P_{\beta-\tau_k}(q_{2k})F_{\alpha_{2k}}(q_{2k-1})\cdots F_{\alpha_1}(q_0)P_\tau(q_0).
\end{align}
Note that $P_\tau(q)$ maps $q\rightarrow q$, and that this matrix product involves the ``string'' of sectors $q_0\rightarrow q_1 \rightarrow \cdots \rightarrow q_{2k}=q_0$, where the final equality is a result of our construction of the CTQMC updates. Often, this string will visit a configuration which is not physical, e.g., that violates the Pauli principal. We say that these ``illegal'' configurations are a part of the zero sector and that any matrix product which would visit sector zero does not survive. Any such string does not contribute to the impurity trace and does not need to be computed. Therefore, we also define a mapping function $s_\alpha(q): q \rightarrow q'$. This allows us to follow the string of visited sectors for every $q_0$ without requiring any costly matrix multiplications, and then compute the trace for the set of $q_0$ which survive the current configuration of operators. 

In general, the operator matrices are of very different sizes and are not necessarily square (although the end result is always square). Still, we can bound the scaling of the new algorithm by assuming every operator matrix is a square matrix of rank $r_{max}$, where $r_{max}$ is the largest dimension of the largest subspace (and operator matrix). If $n_{ss}$ is the number of Hilbert subspaces and all of these subspaces are in the list of surviving $q_0$, the scaling is bounded by $O(k n_{ss}r_{max}^3) $. A lower bound can be defined by noting that if all of the subspaces are of the same size, then $2^n=n_{ss}r_{max}$, such that $O(k n_{ss}r_{max}^3) < O < O(k (2^n)^3 / n_{ss}^3)$. Either bound is dramatically smaller than the naive implementation which does not use Abelian symmetries. Still, the exponential scaling will eventually preclude the evaluation of the local trace in sufficiently large cluster problems unless $n_{ss}=2^n$ (and $r_{max}=1$). (This occurs for a diagonal $t_{ij}$ and Ising-approximated interaction.)

As shown in Fig. \ref{fig:gpu_acceleration_1} and as we will discuss, the size and number of subspaces not only determines the speed of the CTQMC, it also has a large impact on the degree of acceleration a GPU offers. Indeed, GPUs are particularly good at multiplying large matrices when compared to CPUs. 

In addition to this optimization of the local trace computation, we include another major optimization in \ctqmcname{}.
 In general, only a few operators are inserted into the local trace, which might have hundreds of operators. This means that the majority of the subproducts in the trace are left untouched. One can store these subproducts in a data structure, e.g., a balanced binary tree\cite{Gull2011} or a skip-list\cite{semon2014lazy}, to drastically reduce the number of matrix multiplications required. Both of these options offer essentially the same performance improvement, requiring $O[\log(k)]$ matrix multiplications rather than $O(k)$, but the skip-list is much easier to implement.

Furthermore, one can evaluate the bounds of the local trace rather than the trace itself in order to decide whether a proposed configuration will be accepted. (As always, the random number of the Metropolis-Hastings algorithm is drawn first, and this number must be between the lower and upper bounds of the acceptance ratio.) As it is typically much faster to compute these bounds, one can drastically accelerate the CTQMC algorithm by relying upon these bounds to quickly reject proposed moves.  Note that if the proposed move is accepted, the local trace must be computed. However, the acceptance rate is extremely low in real materials at low-temperatures. Therefore, this approach offers a substantial improvement. In the lazy-trace algorithm, the bounds are established by taking the submultiplicative matrix norm of the matrices in the trace. The lazy-trace algorithm utilizes the spectral norm to capture the large variation in the magnitude of the time-evolution operators. 

Now, let us discuss how a GPU can be best used to accelerate these calculations.

\subsubsection{Saturating the GPUs}\label{sec:CUDA}

\begin{figure}
\centering
\includegraphics[scale=1]{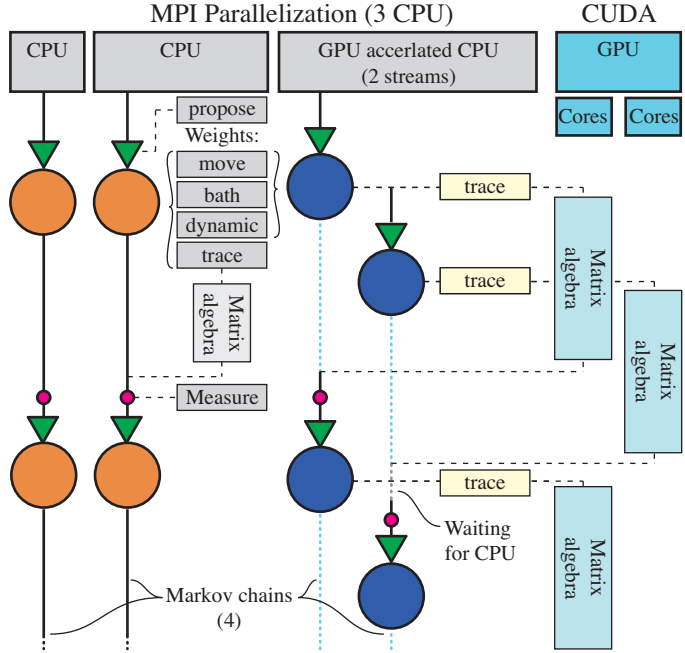}
\caption{Parallelism and acceleration in \ctqmcname. A single Markov chain is created on each MPI image (CPU), with one exception: Each available GPU is paired with a CPU. This GPU accelerated CPU creates a (user specified) number of a Markov chains. While the non-accelerated CPUs handle the local trace computation on their own, the accelerated CPUs hand off the matrix algebra required to the GPU. While waiting for the GPU to handle this algebra, the CPU moves onto the next Markov chain which is not waiting for the GPU to finish its work.  }
\label{fig:parallelism}
\end{figure} 

As we have discussed, the Hilbert space of an impurity quickly becomes prohibitively large, particularly for $f$-shell impurities. Furthermore, this Hilbert space can be decomposed into a block diagonal form according to the symmetries of the local Hamiltonian. The dimension of these blocks is much smaller than the full Hilbert space. In a low-symmetry $f$-shell ($d$-shell) problem, for example, the largest block has a dimension of 858 (28). In a high-symmetry impurity, this number drops to 327 (12).  These matrices are small in comparison to the capabilities of a GPU, even for the high-symmetry $f$-shell impurity. By handling the trace computations for a large number of Markov chains on a single GPU, however, one can still saturate a GPU. There are a number of ways to  accomplish this, most of which are not appropriate for CTQMC, as we will discuss.

Consider a compute node with 7 CPUs and 1 GPU (i.e., a sixth of a Summit node). On this node, we can easily envision each CPU sending instructions to the GPU, such that the GPU has a workload that is seven times larger. However, CUDA will serialize these instructions, and so the GPU will be no more saturated than in the case where one CPU sends instructions to the GPU. The CUDA Multi-Process service (MPS) offers an easy way to parallelize the instructions from multiple CPUs, so that the GPU can simultaneously work on the instructions from all seven CPUs. With MPS, we can indeed have the GPU handle a much larger computational burden. However, MPS essentially partitions the device into seven different units, each of which is paired with a CPU. This means that, 
 each of the seven units must store the operator matrices involved in the local trace, and 
 there will be remainders of each seven units which are not well saturated. 



CUDA streams provide an alternate framework through which to provide the GPU with parallelized instructions. With streams, a CPU divides the instructions it provides to the GPU into multiple streams, and these streams are processed simultaneously by the GPU (if it has the necessary resources). Fortunately, the concept of streams are easily applied to CTQMC: We can associate each CUDA stream with a single Markov chain and have the accelerated CPUs handle multiple of Markov chains. In this framework, a CPU will advance one Markov chain until the local impurity trace must be evaluated, at which point the GPU will begin working on the trace. While the GPU works, the CPU will move on to the next Markov chain, advancing the chain until the trace must be evaluated, at which point the GPU will begin working on that impurity trace (as it simultaneously continues to work on the previous trace). The CPU continues to loop through all of the Markov chains it controls, completing the non-impurity trace work for all Markov chains which are not waiting for the GPU. See a schematic of this process in Fig. \ref{fig:parallelism}.

\begin{figure}
\centering
\includegraphics[scale=1]{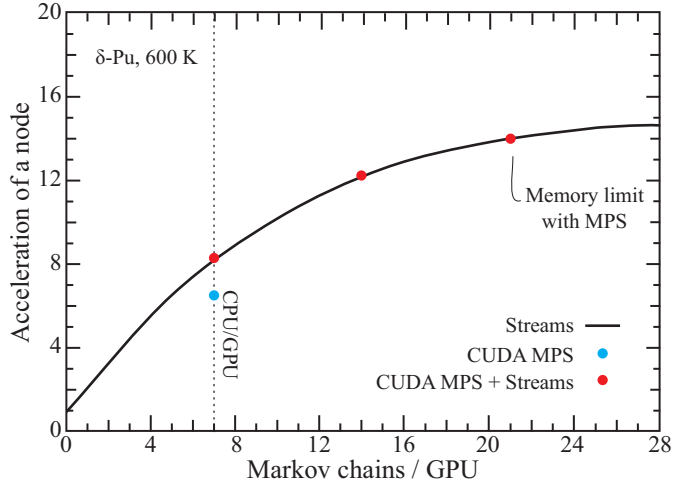}
\caption{ The acceleration offered by various GPU parallelism schemes for variations in the number of Markov chains simulated by each GPU. In Fig. \ref{fig:gpu_acceleration_1}, this plot corresponds to the smallest $f$-shell problem ($n=327$). For the larger $f$-shell problems ($n=858$), MPS cannot be used without running out of memory.  (In contrast with Fig. \ref{fig:gpu_acceleration_1}, acceleration is measured across an entire node.) Tests were run on the supercomputer Summit at ORNL, which features nodes with 42 CPU and 6 GPU (7 CPU / GPU).}
\label{fig:gpu_acceleration_2}
\end{figure}

The global memory of a device is shared between the streams of a single CPU, so that we do consume as much memory as we  would with MPS; and we can dynamically saturate the GPU unit by increasing the number of Markov chains handled by each CPU. Indeed, we can drop MPS entirely and instead have a single CPU handle a large number of simulations. This drastically reduces the global memory requirements.  As shown in Fig. \ref{fig:gpu_acceleration_2}, this approach incurs no penalty in scaling (indicating that the performance is limited by the GPU and CPU-GPU communication, and not by the CPU). Moreover, this figure shows that streams outperform the purely MPS approach, even when the same number of Markov chains are simulated on the GPU.

Our approach is more flexible, portable, and enables strictly better acceleration in all of our tests than a solution which uses MPS. By using it, we are able to achieve a modest acceleration even for $d$-shell impurities, as shown in Fig. \ref{fig:gpu_acceleration_1}. The most impressive acceleration, however, is for the hardest problems. For a complex valued, low-symmetry $f$-shell impurity, a GPU accelerates a single CPU by over 600x on Summit (the six GPUs accelerate the 42 CPU node by over 85x). Without the acceleration offered by GPUs, these problems are prohibitively expensive. For this problem, around 50 node hours are required to gather a reasonable self-energy. For an LDA+DMFT solution, around 1000 node hours would be required. Without using the GPUs, we would require over 600,000 node hours for a single LDA+DMFT run. If we were to try evaluating a more difficult problem, i.e., one with multiple impurities at lower temperatures and with an off-diagonal hybridisation, 10s of millions of node hours would be required for a single LDA+DMFT calculation. If one wanted to gather the vertex functions, the computational cost might exceed the entirety of an INCITE award. Simply put, the GPU acceleration grants access to previously inaccessible regimes, problems, and observables.

This description provides only the broad structure of the GPU-CPU interface. Let us briefly provide more detail on our implementation. 

First, let us overview the GPU computation kernels, the most critical of which are the matrix multiplication and Frobenius norm kernels. The matrix multiplication is handled by the CUTLASS library, from which we design an appropriate double precision matrix-multiplication kernel. The performance of this kernel is very similar to the kernels offered by the cuBLAS libraries. Moreover, it is much more flexible; in the future we hope to achieve an even more dramatic acceleration in the hardest problems by using a mixed-precision matrix-multiplication kernel. (Currently, we are using double-precision throughout the CTQMC. However, the NVIDIA Volta and other HPC devices have dedicated single and half-precision processing units which we do not currently utilize. CUTLASS allows us to target these cores within device code in the future, whereas cuBLAS does not.) Additionally, we implemented CUDA kernel functions for our own norm,reduction, time-evolution, trace, and matrix addition kernels.

Finally, we note that our CPU-GPU structure requires that we only use asynchronous CUDA functions. As memory allocation synchronizes the device, we pre-allocate the entire GPU and develop our own memory allocator which can work asynchronously with that block of pre-allocated memory. One of the additional benefits of our approach is that this allocator catches when the device runs out of memory. When it does, it deletes a stream (and Markov chain) and frees the associated memory. In this way, \ctqmcname{} dynamically finds the optimal number of Markov chains, provided that the number of Markov chains per device (which is specified by the user) is sufficiently high. That is, it is large enough that the device runs out of memory. (Subsequent runs should use the optimized number, as some time will be wasted working on simulations that will be discarded later when the device runs out of memory.)

%% file: Src/Examples.tex
In this section, we will present a results for two examples where we will discuss some best-practices and usage. These examples cover a single-band Hubbard model, which a user can simulate on a modern laptop; and high-symmetry $\delta$-plutonium impurity  which requires access to a cluster and can be accelerated  by a GPU. Here we primarily give the highlights of the more comprehensive tutorials included with the \ctqmcname{} distribution. Note as well that the user guide provided with \ctqmcname{} describes how to install and use \ctqmcname{}, and contains a comprehensive list of the input parameters.

\subsection{Single-Band Hubbard model}
First, we will show how to run a simple example: The impurity of a single-band Hubbard  model. The Hamiltonian for this impurity is
\begin{align}
H=
\sum_{\bm{k}\sigma}\epsilon^{\vphantom{\dagger}}_{\bm{k}}c^\dagger_{\bm{k}\sigma} c^{\vphantom{\dagger}}_{\bm{k}\sigma}
+\epsilon_f\sum_{\sigma} f^\dagger_{\sigma} f^{\vphantom{\dagger}}_{\sigma}
+U n_{f,\uparrow}n_{f,\downarrow}
+\sum_{\bm{k}\sigma} V^{\vphantom{\dagger}}_{\bm{k}}(c^\dagger_{\bm{k}\sigma} f^{\vphantom{\dagger}}_{\sigma} + f^\dagger_{\sigma} c^{\vphantom{\dagger}}_{\bm{k}\sigma}),
\end{align}
where $c^\dagger_{\bm{k}\sigma}$ and $c_{\bm{k}\sigma}$ are the annihilation and creation operators for the non-interacting bath state with wavevector $\bm{k}$, spin $\sigma$, and non-interacting energy $\epsilon_{\bm{k}}$, and $f^\dagger_{\sigma}$ and $f_{\sigma}$ are the annihilation and creation operators for the interacting impurity state with spin $\sigma$ and energy $\epsilon_f$. Here solve this model with three bath states at $\epsilon_{\pm}\pm D/2$ and $\epsilon_0=0$, such that $D=1$ is the bandwidth and our energy scale, and set $U=4D$, $V_{\bm{k}}=V=D$, and $\epsilon_f=-U/2$ (half-filling). 

We choose this simple model and solve it at a moderate temperature, $\beta=10$, for two reasons: First, its possible to run the example on a laptop and replicate these results (aside from the four-point vertex function). Second, exact diagonalization (ED) \cite{Kotliar1996} can quickly provide the exact results for this model. (Note that we get around the issues discussed in \ref{app:EDimpurities} by including a third bath state.) That is, we can compare \ctqmcname{} with the exact solution in order to highlight its various capabilities. 

The input files for this example are provided in \verb|ComCTQMC/examples/hubbard/|, along with a tutorial discussing the choice of parameters and the workflow for running the CTQMC and post-processing codes. Instead of repeating ourselves here, we will instead focus on the results.

\subsubsection{Results}

To provide context for the computational burden associated with these measurements, let us note that these results were computed on a MacBook Pro using 6 of 12 cores on an Intel i9 processor (unless stated otherwise). No GPU was used: For a single-band model, a GPU decelerates the computation as the Hilbert space ($d=2^2=4$) is far too small. Each figure is produced using a different simulation limited to only those worm spaces shown, as it takes substantially more time to gather, e.g., the two-particle vertex than the one-particle vertex (the self-energy). 

\begin{figure}
\centering
\includegraphics[scale=1]{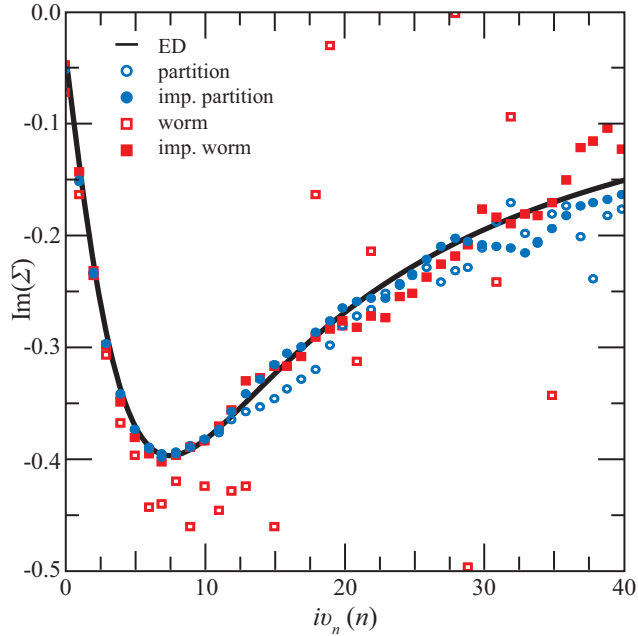}
\caption{The local self energy in the Hubbard model at $\beta=10$. The exact solution is compared with the measurements taken in both worm and partition spaces, with and without the use of improved estimators. The partition space measurements are more accurate, even at moderate expansion orders. (Here, $\langle k \rangle \approx 17$.) The improved estimators significantly improve the results at higher frequencies.  }
\label{fig:SAIM selfen}
\end{figure}

Figure \ref{fig:SAIM selfen} shows the results for the self-energy, the most difficult and important of the one-particle observables to measure. These results were collected in a single 15 minute simulation. As shown, the non-improved partition space measurements are comparable to the improved worm space counterparts, while the improved partition space measurement is the most accurate. In general, as the temperature falls, the partition space measurements increasingly outperform the worm space measurements. As the temperature  rises, the worm space measurements begin to outperform the partition space measurements. Near the atomic limit, the partition space measurements become nearly impossible to converge while the worm measurements converge quickly. 

In light of these observations, it is typically best practice to use the partition space measurement of the self-energy, unless one is near the atomic limit. Not only are the partition space measurements better in the regimes wherein CTQMC is difficult, i.e., at low temperature, but each worm space sampled slows the convergence of all other observables sampled: Consider that the CTQMC must explore $N+1$ configuration spaces, where $N$ is the number of worm spaces. A similar amount of time will be spent in each configuration space due to the volume renormalization using the Wang Landau algorithm and $\eta_{\mathcal{O}}$. (Although the user can specify the fraction of time which should be spent in partition function space.)  The time-to-solution for any observable will therefore increase by a factor of approximately $(N+1)$. 

\begin{figure}
\centering
\includegraphics[scale=1]{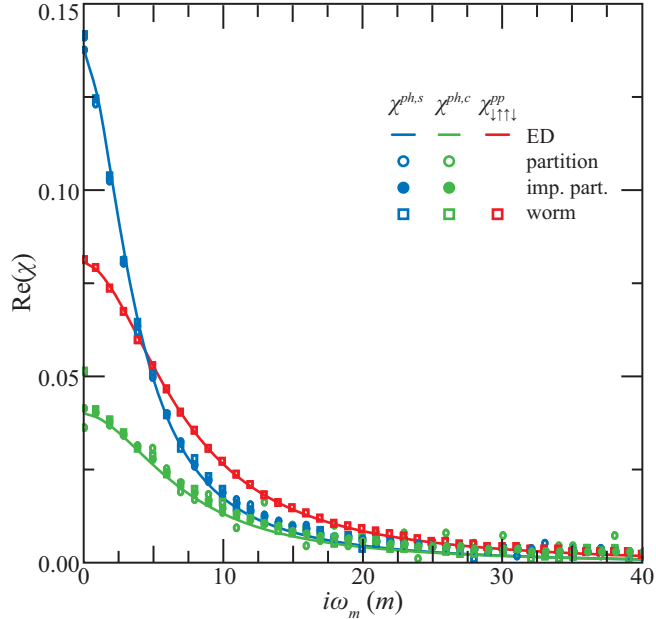}
\caption{The local two-point susceptibilities in the Hubbard model at $\beta=10$. The exact solution is compared with the measurements taken in both worm and partition spaces and with and without improved estimators, when possible. The static two-point susceptibilities in the particle-hole sector are difficult resolve, as are the high-frequency domains of all susceptibilities. The improved estimators help with both regions. }
\label{fig:SAIM susc}
\end{figure}

Figure \ref{fig:SAIM susc} shows the results for the local two-point susceptibilities, measured in a 30 minute simulation. Similar to most objects, the high-frequency region is difficult to resolve. Fortunately, the high-frequency region error does not diverge, as in the self-energy or four-point vertex functions, but simply becomes noisy. In contrast with most objects, the static susceptibilities are also difficult to resolve, and can take the most time to converge. 

As shown, the partition space measurements converge approximately as fast as the worm space measurements. Note that in the Hubbard model all of the susceptibilities in the particle-hole sector can be measured in partition space. As we have discussed, this is not true in multiorbital models: When there are multiple bands, we have many $\chi_{ijkl}\neq 0$ for $i\neq j$ and $k\neq l$. However, if the hybridisation matrix is diagonal, CTQMC can only measure susceptibilities of the form $\chi_{iikk}$ while in partition space. Furthermore, \ctqmcname{} cannot measure any of the particle-particle susceptibilities in partition space, and the partition-space measurements will fail near the atomic limit. Therefore, it is best practice to use the worm space measurements to supplement the partition space measurements (which should be used whenever possible).

\begin{figure}
\centering
\includegraphics[scale=1]{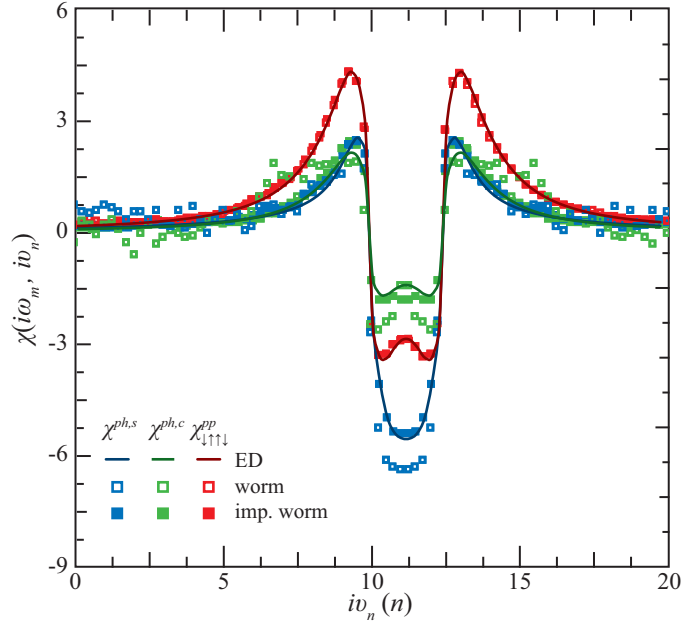}
\caption{The local three-point susceptibilities at $i\omega_{10}$.  Note that partition space measurements are not imiplemented for these objects. Improved estimators greatly increase the accuracy of the measurement, particularly at low ($0<\nu<\omega$) and high frequencies. }
\label{fig:SAIM heden}
\end{figure}

Figures \ref{fig:SAIM heden}  shows the results for the two-particle, three point susceptibilities at a few bosonic frequencies. These results are collected over a 6 hour simulation. Note that these two-time objects are approximately an order of magnitude more difficult to resolve than the one-time Green's functions, at least in this example. Also note that partition space measurements of the  three-point susceptibilities are not implemented. 

Again we see that the improved estimators drastically improve the results, particularly at high frequencies. As with the two-point susceptibilities, the low-frequency region, $0<\nu<\omega$ can also be difficult to resolve, and one should test convergence at both high and low frequencies. The high-frequency Fermionic domain can be resolved using the Legendre basis by setting \verb|basis = "legendre"| in the \verb|hedin| block of the parameter file. However, as with all Legendre basis measurements, one must be careful: The error is systematic rather than stochastic, and it is therefore not as readily apparent. Furthermore, if one is computing the full or irreducible vertex from these three-point susceptibilities, small errors in the susceptibility can lead to large errors in the vertex.

\begin{figure}
\centering
\includegraphics[scale=1]{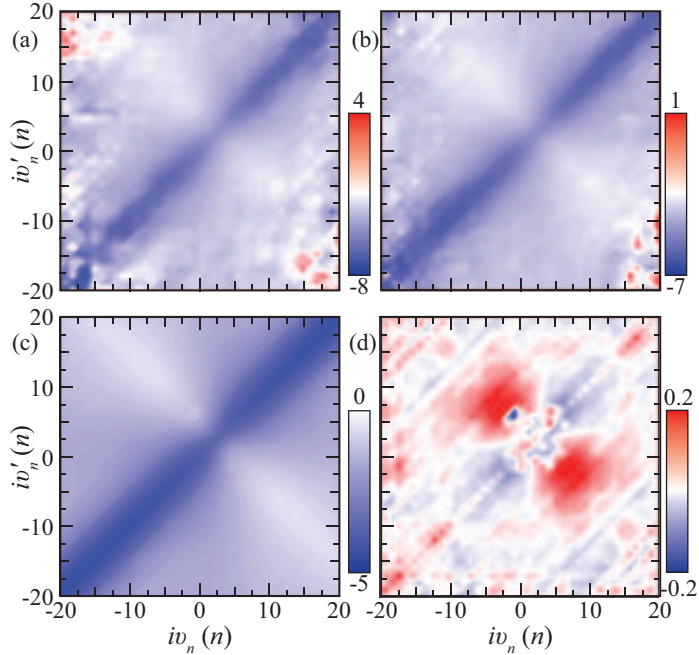}
\caption{The real part of the local full vertex at $i\omega_{5}$ with the (a) Fourier and (b) mixed Legendre-Fourier basis, and (c) using the vertex asymptotics. (d) The difference between the (c) and the exact, ED result. Improved estimators are used for every observable but the local two-point susceptibilities and results are accumulated across 20,000 CPU hours. Even for this simple problem, simulated for a comparatively long time and using improved estimators, the asymptotics are crucial when resolving the vertex at an high frequencies. }
\label{fig:SAIM vertex}
\end{figure}

Figure\ref{fig:SAIM vertex} shows the results for the full four-point vertex in the charge channel. (The full vertex is the four point susceptibility with its legs amputated, $F_{ijkl}=\chi_{ijkl}/G_iG_jG_kG_l$.) The four point vertices require by far the most computation time for a good estimate. Here we present results using improved estimators accumulated across simulations requiring nearly 2,000 CPU hours. Still, the error in the vertex at high frequencies is unacceptable, even in the Legendre basis. By using the vertex asymptotics, however, we can very nearly recover the exact (ED) result at arbitrary frequencies. 

Note that in order to compute the vertex asypmtotics, we must measure the two and three point susceptibilities in both particle-hole and particle-particle channels. Therefore, one must sample six configuration spaces (instead of just two). In some sense, this magnifies the computational burden. However, the benefits greatly outweigh the costs, particularly if one requires the high-frequency components, as shown in Fig. \ref{fig:SAIM vertex}.

\subsection{$\delta$-Plutonium}

Now, let us move on to a real material: $\delta$-Plutonium.  $\delta$-Plutonium is a high-temperature form of Plutonium with only one atom in the unit cell. This is one of the easiest $f$-shell problems for CTQMC to solve. Here we will discuss the final CTQMC simulation of a converged LDA+DMFT relativistic run using ComDMFT\cite{Choi2019}. In this case, ComDMFT uses the the default one-particle spin-orbit coupled basis set described in \ref{app:basis}, and discards the off-diagonal elements of the hybridisation and one-body Hamiltonian. (This leads to many more symmetries than might be expected in a relativistic, $f$-shell impurity, e.g., $J_z$ remains a good quantum number.)  We use the following parameterization for the local two-body interaction and the nominal double counting scheme:  Hubbard interaction $U=4.5$, Hund interaction $J=0.512$, and nominal occupancy of the impurity orbitals $N_0=5$. The results of the DFT+DMFT simulation are sensitive to this parameterisation and the double counting model. See, for example, Ref. \cite{Zhu_2007}. However, it is not our intention to present a treatise on the most accurate simulation of $\delta$-Plutonium within DFT+DMFT. Instead, we would like to restrict our analysis to the usage, accuracy, and computational cost of the CTQMC simulation. Therefore, we will leave a more in-depth discussion of the double counting for a later study and simply use a parameterization that lets us compare our results to a recent publication.

\subsubsection{Results}

\begin{figure}
\centering
\includegraphics[scale=1]{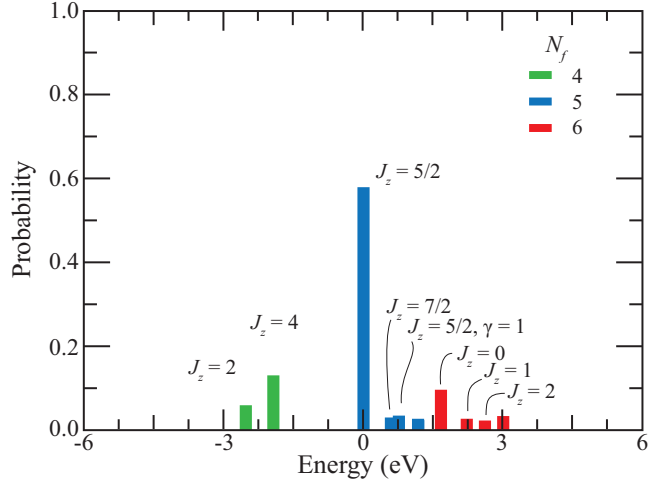}
\caption{Valence histogram of $\delta$-Pu at 600 K. Valence fluctuations between the $N=5$ ground state and the $N=6$ excited states help to explain the anomalous properties of $\delta$-Pu.\cite{Shim2007plutonium} ComCTQMC can generate histograms for any quantum number which commutes with the local Hamiltonian, which can be crucial in efforts to describe the correlated physics of a strongly-correlated material.  }\label{fig:dpu:histogram}
\end{figure}

\begin{figure}
\centering
\includegraphics[scale=1]{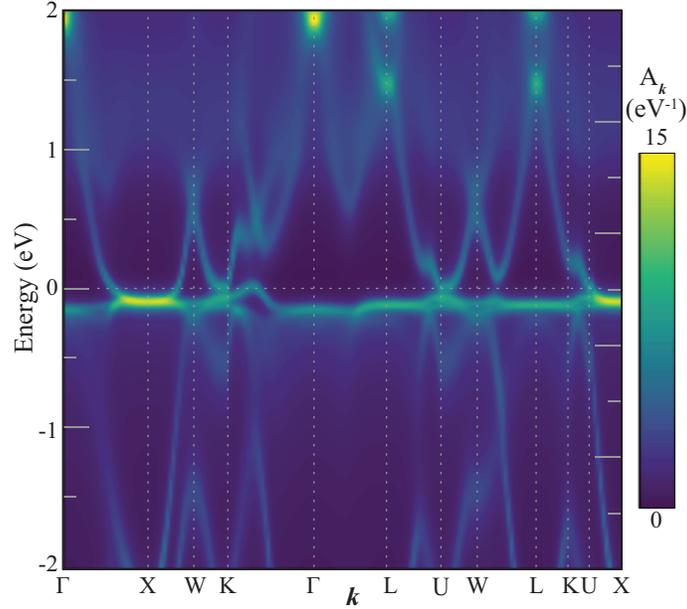}
\caption{The momentum resolved spectral function, $A_{\bm{k}}$, of $\delta$-Pu at 600 K along the high-symmetry lines. The spectrum features Hubbard and quasiparticle bands, with a Kondo peak just below the Fermi level. These results compare favorably with a recently published study on $\delta$-Pu,  while  requiring drastically less computer time.\cite{Tutchton2020}}\label{fig:dpu:spectral}
\end{figure}

In Fig. \ref{fig:dpu:histogram}, we present the valence histogram resulting from these calculations. As has been discussed in the literature, $\delta$-Pu exhibits significant valence fluctuations between the ground-state valence configuration, $N=5$ and $J_z=5/2$, and a number of other configurations. Primarily, it fluctuates to a number of higher energy, $N=6$ states.  This differs substantially from, e.g., curium, which contains a single peak in the histogram (the ground state) with a probability near unity \cite{Shim2007plutonium}. This valence histograms helps to explain the anomalous behavior of Plutonium, e.g., its  lack of magnetism, and the ability to quickly and easily produce such a histogram is enormously useful in the effort  to understand various strongly correlated materials. The user guide for \ctqmcname{} goes into the calculation of this histogram or histograms like it using \ctqmcname{}.

In Fig. \ref{fig:dpu:spectral}, we present the spectral function of $\delta$-Pu. To produce the spectral function, we must first analytically continue the imaginary domain self-energy produced by \ctqmcname{}. ComCTQMC does not have a built-in analytical continuation program, as a number of well-developed, open-source  analytical continuation codes have already been developed and published. Here we use the maximum entropy code bundled into  the extended-DMFT package (EDMFT)\cite{EDMFT}. As shown in Fig. \ref{fig:dpu:spectral}, the spectral function of $\delta$-Pu is dominated by a Kondo peak near the Fermi level (which does not appear in pure LDA or GGA calculations).\cite{Savrasov2001} 

To get these results, the LDA+DMFT simulation was run for 20 iterations on 50 nodes on Summit at ORNL. Each iteration required 10 minutes of measurement by ComCTQMC, with the first iteration given another 5 minutes of thermalisation. An additional 20 minutes of measurement time was allocated to the final two iterations for the purpose of gathering more refined results for the analytical continuation. In total, around 200 node hours were spent on the solution. This is a shockingly small investment into a problem that has, until now, been prohibitively expensive for most researchers. For comparison, the first CTQMC simulations of Pu required ``massive computer resources", i.e., a major allocation on a leadership-class facility \cite{Marianetti2008}. Suddenly, research into actinide systems is not only possible, but it can also be inexpensive in many cases. Furthermore, those problems which were previously impossible without additional approximation are now possible.

\begin{figure}
\centering
\includegraphics[scale=1]{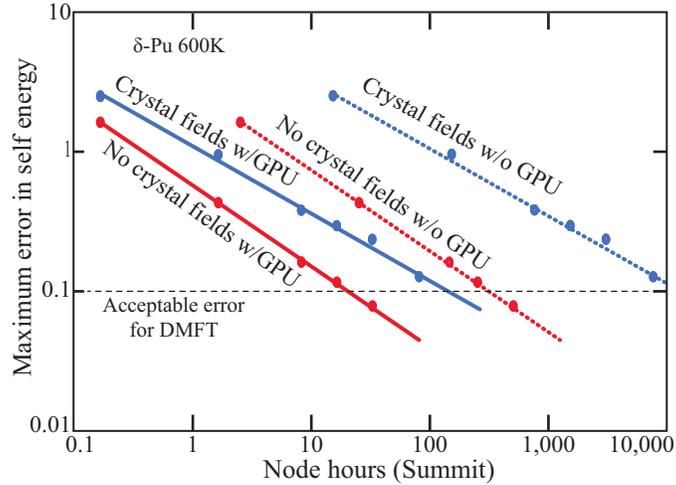}
\caption{Relative error in the self-energy at 10 meV, i.e., the maximum error in the DMFT loop, as the number of node hours allocated to \ctqmcname{} increases.  Without crystal fields, the GPU acceleration transforms a DFT+DMFT simulation (~25 times as expensive as a single \ctqmcname{} run) from moderately expensive to inexpensive. With crystal fields, the GPU acceleration transforms the DFT+DMFT simulation from nearly impossible to easily doable. Indeed, the GPU acceleration will allow researchers to extend the boundaries in temperature and unit-cell size and simulate, e.g., not only the high-temperature actinide systems like $\delta$-Pu but also the low temperature systems like $\alpha$-Pu.
}
\label{fig:time-to-solution}
\end{figure}

Consider, for example an LDA+DMFT simulation of $\delta$-Pu with crystal field effects. This simulation would require approximately 250,000 node hours on Summit without GPU acceleration, as shown in Fig. \ref{fig:time-to-solution}. While possible, this would represent a massive computational effort. With GPU acceleration, we reduce the computational burden by two orders of magnitude! Suddenly, it is relatively inexpensive to compute the one-particle observables in the presence of crystal fields. These fields should have a pronounced impact on observables which exist at very low energy scales (e.g., the Fermi surface) or on many observables in materials with less symmetry than $\delta$-Pu (e.g., the lower-temperature Plutonium allotropes), and so it is important that \ctqmcname{} allows us to include them without requiring exorbitant  expense. As these materials are already enormously expensive to simulate (due the large number of impurities in the unit cell and the low temperatures at  which  they are stable), the GPU acceleration achieved by \ctqmcname{} will be necessary to accurately simulate them within DMFT. \footnote{At low temperatures, CTQMC encounters the sign problem, and one cannot overcome the sign problem by brute force. Still, the sign problem is not what currently prevents researchers from simulating lower symmetry actinide systems like $\alpha-Pu$ with DFT+DMFT, particularly if they include the crystal fields.} 

Similarly, gathering two-particle observables like the susceptibilities or vertex functions is extraordinarily costly, particularly if they must be analytically continued. It requires approximately an order of magnitude more computer time to measure the local susceptibility $\chi^{(2,ph)}_{susc}$ than it does to measure the self-energy (with the same degree of relative error).\footnote{\ctqmcname{} estimates the absolute error if the user directs it to do so. This is done by comparing the estimators accumulated on each MPI rank using the jackknife method. Alternately, it is done by comparing sequential, serial runs if the user does not have MPI.} To get a good analytical continuation, it requires at least two-orders of magnitude more computer time than the entire LDA+DMFT simulation. It is crucial, then, that the GPU offers one-to-two orders of magnitude reduction in the computational burden, particularly as researchers look towards the two-particle vertex functions.

%% file: Src/Appendices.tex
\section{Basis functions}
\label{app:basis}

In general, one does not need to know the form of the one-particle basis to run \ctqmcname. However, \ctqmcname{} provides Slater-Condon and Kanamori parameterisations of the static two-body interaction tensor. In order to use these parameterisations, it is necessary to know this basis set. For this  purpose, \ctqmcname{} implements three basis sets. A generic  basis  set,  which is all that is required for the Kanamori interaction; and the product and spin-coupled basis sets, which provide the detail required for the Slater-Condon interaction.

The product  basis set is defined here as
\begin{align}
|l,m,\sigma\rangle = Y_{l,m} \otimes \sigma,
\end{align}
where  $\sigma$ is the spin and $Y_{l,m}$ are the real spherical harmonics, which are defined as
\begin{equation}
\begin{aligned}Y_{\ell, m}:= {
\begin{cases} {i  {\sqrt {2}}}\left(Y_{\ell }^{m}-(-1)^{m}\,Y_{\ell }^{-m}\right)&{\text{if}}\ m<0\\ 
Y_{\ell }^{0}&{\text{if}}\ m=0\\
{1  {\sqrt {2}}}\left(Y_{\ell }^{-m}+(-1)^{m}\,Y_{\ell }^{m}\right)&{\text{if}}\ m>0\end{cases}}
\end{aligned}
\end{equation}
in terms of the complex spherical harmonics $Y_l^m$.

The coupled basis set is defined here as 
\begin{align}
|j,m_j\rangle = \sum_{m\sigma} Y_{l}^m\otimes \sigma \langle \ell,m;\frac{1}{2},\sigma | j,m_j\rangle,
\end{align}
where $\langle \ell,m;\frac{1}{2},\sigma | j,m_j\rangle$ are the Clebsch-Gordan coefficients.

The generic, product, and coupled basis are enumerated as 
\begin{align}
&|1,\downarrow\rangle, ... , |n,\downarrow\rangle, |1,\uparrow\rangle, ... , |n,\uparrow\rangle \\
&|\ell,-\ell,\downarrow \rangle , \,\, |\ell,-\ell + 1,\downarrow \rangle , \dots, \,\, |\ell,\ell,\downarrow \rangle, \,\, |\ell,-\ell,\uparrow \rangle, \,\, |\ell,-\ell + 1,\uparrow \rangle,  \dots, \,\, |\ell, \ell,\uparrow \rangle 
\label{equ:enumProduct}\\
&|\ell-\frac{1}{2}, -\ell + \frac{1}{2}\rangle, \dots,\,\,  | \ell-\frac{1}{2}, \ell-\frac{1}{2}\rangle,\,\,  | \ell+\frac{1}{2}, -\ell-\frac{1}{2}\rangle, \dots,\,\,  | \ell+\frac{1}{2} , \ell+\frac{1}{2}\rangle. \label{equ:enumCoupled}
\end{align}

\section{Metropolis-Hasting algorithm}\label{sec:metropolis-hasting}
A Markov chain is a random walk $\boldsymbol x_0 \longrightarrow \boldsymbol x_1 \longrightarrow  \cdots$ in the sample space, characterized by the transition probability $P(\boldsymbol x_{n+1}|\boldsymbol x_n)$ between two consecutive samples. Given the probability distribution $p_{n}(\boldsymbol x_n)$ at the step $n$, the probability distribution at step $n+1$ reads
\begin{equation}
\label{equ:MarkowConvergence}
p_{n+1}(\boldsymbol x_{n+1}) = \sum_{\boldsymbol x_n} P(\boldsymbol x_{n+1}|\boldsymbol x_n)p_n(\boldsymbol x_n), 
\end{equation}
and if the Markov chain converges, the stationary distribution $p := p_n$ $(n\rightarrow \infty)$ satisfies global balance
\begin{equation} 
\label{equ:globalBalance}
p(\boldsymbol y) = \sum_{\boldsymbol x} P(\boldsymbol y|\boldsymbol x)p(\boldsymbol x).
\end{equation}
The goal is now to find such a transition probability $P$ for a given target distribution $p$. A sufficient (but not necessary condition) for global balance is detailed balance
\begin{equation}
P(\boldsymbol y|\boldsymbol x)p(\boldsymbol x) = P(\boldsymbol x|\boldsymbol y)p(\boldsymbol y).
\end{equation}
Since the probability of going from state $\boldsymbol x$ to any state $\boldsymbol y$ is one, this implies
\begin{equation}
\sum_x P(\boldsymbol y|\boldsymbol x)p(\boldsymbol x) = \sum_{\boldsymbol x} P(\boldsymbol x|\boldsymbol y)p(\boldsymbol y) = p(\boldsymbol y).
\end{equation}
A general solution of detailed balance is provided by the Metropolis-Hasting algorithm. A new configuration $\boldsymbol y$ is proposed with probability $P_{\text{prop}}( \boldsymbol y | \boldsymbol x)$, and accepted with probability $P_\text{acc}(\boldsymbol y | \boldsymbol x)$. If the proposed configuration $\boldsymbol y$ is rejected, the old configuration $\boldsymbol x$ is used again. With $\boldsymbol y \ne \boldsymbol x$, the detailed balance condition for this transition probability $P(\boldsymbol y | \boldsymbol x)=P_\text{acc}(\boldsymbol y | \boldsymbol x)P_{\text{prop}}( \boldsymbol y | \boldsymbol x)$ implies
\begin{equation}
\label{equ:MHratio}
\frac{P_{\text{acc}}(\boldsymbol y | \boldsymbol x)}{P_\text{acc}(\boldsymbol x | \boldsymbol y)} = \frac{P_\text{prop}(\boldsymbol x | \boldsymbol y)p(\boldsymbol y)}{P_\text{prop}(\boldsymbol y | \boldsymbol x)p(\boldsymbol x)} =: R(\boldsymbol y, \boldsymbol x),
\end{equation}
where $R(\boldsymbol y, \boldsymbol x)$ is the acceptance ratio, and is satisfied by Metropolis-Hasting choice
\begin{equation}
\label{equ:decide}
P_\text{acc}(\boldsymbol y | \boldsymbol x) = \min (1, R(\boldsymbol y, \boldsymbol x)).
\end{equation}
For $\boldsymbol x = \boldsymbol y$, detailed balance is trivially satisfied. Notice that this algorithm 
allows to get samples of a distribution which is only known upon a constant factor.

\section{Limitation of the Green function estimator in partition function space}
\label{app:EDimpurities}
In this appendix, we show that the estimator for the Green function in partition space in Eq.~\ref{equ:PGreenEstimator} can have limitations if the bath has only very few orbitals. While this is not relevant for DMFT calculations, it can be useful when benchmarking a CTQMC solver with ED. 

To start with, we consider a toy impurity model with two impurity orbitals and two bath orbitals. The Hamiltonian reads
\begin{equation}
\label{equ:ProofHamiltonian}
H=\overbrace{\sum_{i,j=1,2} \hat{c}^\dagger_i t_{ij} \hat{c}_j}^{H_\text{loc}} + \overbrace{\sum_{i=1,2}\hat{f}^\dagger_i \hat{c}_i + \hat{c}_i^\dagger \hat{f}_i}^{H_\text{hyb}} + \overbrace{\sum_{i=1,2} \hat{f}^\dagger_i \hat{f}_i}^{H_\text{bath}},
\end{equation} 
where the off-diagonal elements of the hopping matrix $t_{ij}$ are non-zero. 

The partition estimator for the Green function is based on the assumption that the Green function weight $w(\mathcal{C}, \hat{c}(\tau') \hat{c}^\dagger(\tau))$ vanishes whenever the partition function weight $w(\mathcal C + \tau\tau')$ vanishes. This assumption is violated for the configuration $\mathcal C = \hat{c}_1^\dagger(\tilde{\tau})\hat{c}_1(\tilde{\tau}') $ and the Green function entry $\hat{c}_1(\tau') \hat{c}_1^\dagger(\tau)$ with $\tilde{\tau} > \tilde{\tau}' > \tau' > \tau$. The Green function weight 
\begin{equation}
\label{equ:ProofGreenWeight}
w(\mathcal{C}, \hat{c}(\tau') \hat{c}^\dagger(\tau)) =Z_\text{loc}\overbrace{\langle \hat{c}^\dagger(\tilde{\tau}) \hat{c}(\tilde{\tau}') \hat{c}(\tau') \hat{c}^\dagger(\tau) \rangle_\text{loc}}^{\ne 0} \times \overbrace{\langle \hat{f}(\tilde{\tau})\hat{f}^\dagger(\tilde{\tau}') \rangle_\text{bath}}^{\ne 0},
\end{equation}
is non-zero. Here the orbital indices are omitted, and 
\begin{align}
\langle \circ \rangle_{\text{loc/bath}}=Z_\text{loc/bath}\mathrm{Tr}e^{-\beta H_\text{loc/bath}}\circ.
\end{align}
 The partition function weight  
\begin{equation}
\label{equ:ProofPartitionWeight}
w(\mathcal C + \tau\tau') = Z_\text{loc}\overbrace{\langle \hat{c}^\dagger(\tilde{\tau}) \hat{c}(\tilde{\tau}') \hat{c}(\tau') \hat{c}^\dagger(\tau) \rangle_\text{loc}}^{\ne 0} \times \overbrace{\langle \hat{f}(\tilde{\tau}) \hat{f}^\dagger(\tilde{\tau}') \hat{f}^\dagger(\tau') \hat{f}(\tau) \rangle_\text{bath}}^{=0}
\end{equation}
however is zero. This is because two bath creation (annihilation) operators lie next to each other and, as opposed to the fermions on the impurity, the fermions in the bath can not hop between the two orbitals. 
It is easily seen that contribution to the Green function as in Eq.~\ref{equ:ProofGreenWeight}, which can not be sampled by the partition space Green estimator, appear at arbitrary expansion orders.

In Eq.~\ref{equ:ProofHamiltonian} we have chosen the simplest Hamiltonian which violates the assumption for the Green function estimator in partition space that $w(\mathcal{C}, \hat{c}(\tau') \hat{c}^\dagger(\tau))$ vanishes whenever $w(\mathcal C + \tau\tau')$ vanishes. The same happens if the one body part of the impurity Hamiltonian $H_\text{loc}$ is diagonal, but we add a Kanamori interaction which shuffles the fermions on the impurity through spin-flip and pair-hopping processes.

\section{CTQMC Move optimizations}\label{sec:CTQMCMoveOptimizations}

While the trace calculation is an unavoidable bottleneck in CT-HYB calculations, a substantial amount of computation can be avoided by only proposing moves which have a chance to be accepted. That is, we often know \textit{a priori} whether a move will be rejected, without having to do an expensive local trace computation in addition to the moderately burdensome hybridisation and dynamical interaction computations. 

\begin{figure}
\centering
\includegraphics[scale=1]{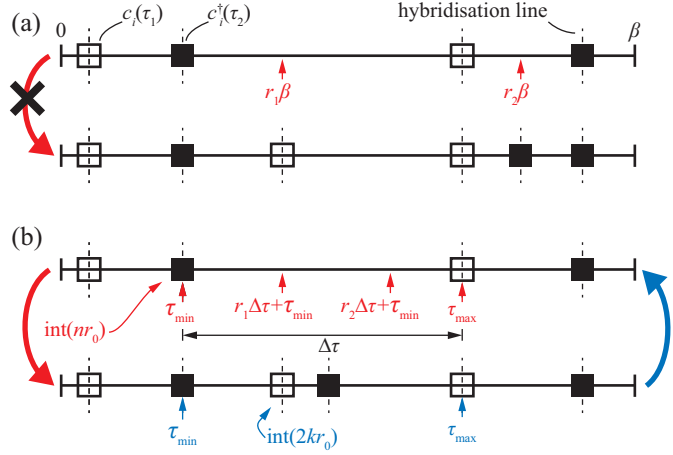}
\caption{ The (a) standard and (b) Pauli-aware pair insertion and removal moves, where $r_i$ are the random numbers required for the move and $n$ is the number of operators in the current configuration. In the standard algorithm, many randomly proposed moves will be rejected because they generate unphysical configurations. In the Pauli-aware algorithm, the acceptance rate is much higher, particularly at high expansion orders.  }
\label{fig:CSCmoves}
\end{figure}

Consider the pair insertion move. In the literature, this move is typically made by generating two random times for the creation and annihilation operator, computing the Metropolis-Hastings acceptance rate, and then deciding to accept or reject the proposed move\cite{Gull2011}, as shown in Fig. \ref{fig:CSCmoves}(a). Often, however, the proposed move will violate the Pauli-exclusion principle, and we have wasted effort computing the acceptance rate. In \ctqmcname, we only propose moves that satisfy the Pauli exclusion principle. In order to accomplish this, we must take extra care with the pair insertion, removal, and worm operator shift moves. Let us briefly outline how we propose the Pauli-aware moves.

\textbf{Pair insertion}: First, one selects the flavor of the creation and annihilation operator, $i$ and $j$. Then, we select a random operator from the list of operators of flavor $i$ and $j$ in the current configuration. Next, we generate two random times which lie on the imaginary time axis between this operator and the next operator in the time-ordered list. (If there are no operators in the current configuration of flavor $i$ and $j$, then one simply chooses two random times as in the typical, Pauli-unaware algorithm.) As in the Pauli-unaware algorithm, one must attach a weight to the Metropolis-Hastings acceptance rate in order to maintain the detailed balance. We call that weight the configuration weight, $w_{\mathrm{config}}$. For this Pauli-aware pair insertion move, the configuration weight is
\begin{align}
w_{\mathrm{config}}=\Delta \tau \frac{N}{(N+2)},
\end{align} 
where $\Delta\tau$ is the distance between the pre-existing operators which bound the randomly select times for the newly proposed pair, and $N$ is the number of operators of flavor $i$ and $j$. Note the imaginary time axis is a circle of circumference $\beta$, so that one must take care with the distance calculation and concept of the ``next'' operator. This move is illustrated in Fig. \ref{fig:CSCmoves}(b).

\textbf{Pair removal}: This move proceeds much like the pair insertion move. That is, we select the flavors of the creation and annihilation operators and then select a random operator from the list of operators with these flavors. Then, we select the next operator in the time-ordered list. Finally, we compute the distance between between these pairs along the imaginary time axis, a circle. The resulting configuration weight is
\begin{align}
w_{\mathrm{config}} = \frac{1}{\Delta\tau}\frac{N}{N-2}.
\end{align}
This move is illustrated in Fig. \ref{fig:CSCmoves}(b).

\begin{figure}
\centering
\includegraphics[scale=1]{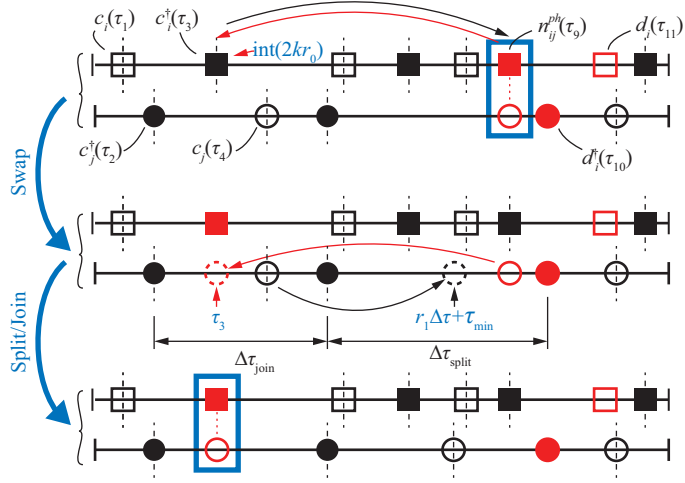}
\caption{ The algorithm for moving a local bilinear. In order to generate physical configurations, the bilinear is moved throughout the configuration by swapping one of its operators with a hybridized (configuration space) operator forming a new bilinear at this location, and splitting apart the old bilinear. The operators used for the split and join portions of this move are aware of the Pauli exclusion principle, greatly improving the acceptance rate at large expansion orders.
   }
\label{fig:SwapSplitJoin}
\end{figure}

\textbf{Bilinear swap}: In general, one must swap the location of worm and partition operators in order for the exploration of worm configuration spaces to be ergodic. This is a relatively simple process when one does not  need to swap a bilinear operator: Pauli exclusion is never violated and one only needs to compute the change in the hybridisation.\cite{Gunacker2015} When one swaps one of the operators in a bilinear, however, the other operator in that bilinear must be dealt with carefully. If one simply shifts it, Pauli exclusion will be violated quite often, which might cause ergodicity problems. Moreover, if both operators in a bilinear are swapped, a newly hybridized bilinear is left in place. This hybridized bilinear is relatively unphysical and will therefore typically be rejected by the Metropolis-Hastings algorithm. Again, ergodicity issues arise. Therefore, we must carefully design a smart bilinear swap move. In \ctqmcname, we implement the bilinear swap using a combination swap, join, and split moves in order to avoid the problems outlined above. This move is illustrated in Fig. \ref{fig:SwapSplitJoin}.

As in the typical swap move, the join-swap-split move for a bilinear $n^{(l)}_{ij}$ begins by selecting a random hybridized operator of flavor $i$ ($j$) from the current configuration. This time, $\tau_{\mathrm{join}}$ is where the new bilinear will be placed, and these are the operators which will be swapped (hybridized $\leftrightarrow$ local). 
Then, one finds the operators of flavor $j$ ($i$) which bracket the current and new bilinears. A random time is selected within the operators bracketing the current bilinear, $\tau_{\mathrm{split}}$, and we select the operator of the same type (creation or annihilation) from the operators bracketing the new bilinear. The hybridized operator is moved to $\tau_{\mathrm{split}}$ and the local operator is moved to $\tau_{join}$
The configuration weight associated with this move is 
\begin{align}
w_{\mathrm{config}} = \frac{\Delta \tau_{\mathrm{split}}}{\Delta\tau_{\mathrm{join}}},
\end{align}
where $\Delta\tau_{\mathrm{split}}$ and $\Delta\tau_{\mathrm{join}}$ are the distances between the operators bracketing the original and new bilinears, respectively.

\section{Symmetries of the two-particle correlators}\label{app:susceptibilities}

 In general, accurate measurement of the two-particle correlators is much more time consuming than the accurate measurement of the one-particle correlators. In \ctqmcname, we implement a number of symmetries during post-processing which help to improve the accuracy of the measurements. The symmetries implemented in \ctqmcname{} are the Hermetian and Fermionic symmetries, when possible. We list these symmetries in Table \ref{tab:symmetries}
\begin{table}
\caption{Symmetries implemented in \ctqmcname}\label{tab:symmetries}
\begin{center}
\begin{tabular}{c c c}
 \multicolumn{3}{c}{\textbf{Observable}} \\
 $G^{(ph)}_{ijkl}(i\omega)$ & $G^{(pp)}_{ijkl}(i\omega)$ & $G^{(ph)}_{ijkl}(i\nu,i\omega)$  \vspace{1mm} \\
\hline\vspace{2mm}
{$\!\begin{aligned}
&G^{(ph)}_{jilk}(i\omega) \\
&G^{(ph)}_{lkji}(-i\omega) \\
&G^{(ph)}_{klij}(-i\omega)
 \end{aligned}$}
& 
{$\!\begin{aligned}
&G^{(pp)}_{jikl}(i\omega) \\
&G^{(pp)}_{ijlk}(i\omega) \\
&G^{(pp)}_{jilk}(i\omega)
 \end{aligned}$}
&
{$\!\begin{aligned}
&G^{(ph)}_{jilk}(i\omega-i\nu,i\omega) 
 \end{aligned}$} \\
 &&\\
 \multicolumn{3}{c}{\textbf{Observable}} \\
 $G^{(pp)}_{ijkl}(i\nu,i\omega)$  &  \multicolumn{2}{c}{$G^{(ph)}_{ijkl}(i\nu,i\nu',i\omega)$} \vspace{1mm} \\
\hline\vspace{2mm}
{$\!\begin{aligned}
&G^{(pp)}_{ijlk}(i\nu,i\omega) \\
&G^{(pp)}_{jikl}(i\omega-i\nu,i\omega) \\
&G^{(pp)}_{jilk}(i\omega-i\nu,i\omega)
 \end{aligned}$}
 &
 \multicolumn{2}{c}{
{$\!\begin{aligned}
 &G^{(ph)}_{kjil}(i\nu'-i\omega,i\nu',i\nu'-i\nu) \\
 &G^{(ph)}_{ilkj}(i\nu-i\omega,i\nu,i\nu-i\nu') \\
 &G^{(ph)}_{lkij}(i\nu'-i\omega,i\nu-i\omega,i\omega) 
 \end{aligned}$}
 }
 \end{tabular}
 \end{center}
\end{table} 
 
If a symmetric component specified above is not measured (i.e., it is outside the frequency box defined by the user supplied cutoff frequencies), that symmetry is not applied at that frequency. That is, these symmetries are applied whenever possible and ignored otherwise. These symmetries are also implemented for the improved estimators.
 
 \section{Discussion of Quantitative Scaling and Acceleration Results}\label{app:Summit}

In general, the scaling and acceleration of \ctqmcname{} will depend on not only the parameterization of the CTQMC, but it will also depend upon the architecture of the computational resource, the compilers and libraries used, and the flags set when compiling these libraries and \ctqmcname{} itself. Of particular importance is the relative power of the collection of GPUs and CPUs on a node: Summit, for example, contains a few powerful GPUs and fairly slow (but numerous) CPU cores. To some extent, this leads to the drastic acceleration of a single core shown in Fig. \ref{fig:gpu_acceleration_1}. 

Another factor which can have a noticeable effect is the linear algebra library: If the matrix multiplication is well optimized for the CPU architecture, as in the Intel MKL  (math kernel library), GPU matrix multiplication loses some of its advantage, particularly for moderate-to-small sized matrices.  Scaling across nodes depends primarily on the MPI library and communication infrastructure of the resource. However, ideal weak scaling is almost always achievable if sufficiently long measurement times are employed.  

The results presented here (for scaling and acceleration) used the following compilers and libraries: (C++ compiler)  G++ version 7.4.0 with optimization level O3, (CUDA library and compiler) CUDA version 10.1.243, (MPI library) Spectrum-MPI version 10.3.1.2-20200121, and  (linear algebra library) ESSL version 6.1.0-2. The results were accumulated from Summit runs performed throughout March 2020.